\documentclass[10pt,fleqn]{article}
\usepackage{amsmath}
\usepackage{amssymb}

\usepackage{graphicx}

\usepackage{cite}

\usepackage{color} 
\usepackage{soul}

\usepackage[labelfont=bf,labelsep=period,justification=raggedright]{caption}

\makeatletter
\renewcommand{\@biblabel}[1]{\quad#1.}
\makeatother

\date{}

\newcommand{\be}{\begin{equation}}
\newcommand{\ee}{\end{equation}}
\newcommand{\bea}{\begin{eqnarray}}
\newcommand{\eea}{\end{eqnarray}}

\def\<{{\langle}}
\def\>{{\rangle}}
\newcommand{\sectionlarge}[1]{\section*{\LARGE #1}}
\newcommand{\subsectionmedium}[1]{\subsection*{\large #1}}
\newcommand{\subsubsectionsmall}[1]{\subsubsection*{\small #1}}

\begin{document}

\begin{flushleft}
{\Large
\textbf{Intrinsic noise of microRNA-regulated genes and the ceRNA hypothesis}
}
\\
Javad Noorbakhsh$^{1}$, 
Alex H. Lang$^{1}$, 
Pankaj Mehta$^{1,\ast}$
\\
\bf{1} Physics Department, Boston University, Boston, MA, USA
\\
$\ast$ E-mail: pankajm@bu.edu
\end{flushleft}

\sectionlarge{Abstract}
MicroRNAs are small noncoding RNAs that regulate genes post-transciptionally by binding and degrading target eukaryotic mRNAs. We use a quantitative model to study gene regulation by {inhibitory} microRNAs and compare it to gene regulation by prokaryotic small non-coding RNAs (sRNAs). Our model uses a combination of analytic techniques as well as computational simulations to calculate the mean-expression and noise profiles of genes regulated by both microRNAs and sRNAs. We find that despite very different molecular machinery and modes of action (catalytic vs stoichiometric), the mean expression levels and noise profiles of microRNA-regulated genes are almost identical to genes regulated by prokaryotic sRNAs. This behavior is extremely robust and persists across a wide range of biologically relevant parameters. We extend our model to study crosstalk between multiple mRNAs that are regulated by a single microRNA and show that noise is a sensitive measure of microRNA-mediated interaction between mRNAs. We conclude by discussing possible experimental strategies for uncovering the microRNA-mRNA interactions and testing the competing endogenous RNA (ceRNA) hypothesis.


\sectionlarge{Introduction}
MicroRNAs are short sequences of RNA ($\simeq$22 base pairs) that post-transcriptionally regulate gene expression in eukaryotes by destabilizing target mRNAs \cite{Ambros2004,Reinhart2002}. Since their discovery almost two decades ago \cite{Lee1993}, there has been a steady increase in the number of discovered microRNAs. MicroRNAs play an important role in many biological processes, including animal development \cite{Alvarez-Garcia2005,Miska2005}, tumor suppression \cite{Caldas2005,Shenouda2009}, synaptic development \cite{OldeLoohuis2012,Schratt2009}, programmed cell death \cite{Fuchs2011,Baehrecke2003}, and hematopoietic cell fate decisions\cite{Havelange2010,Chen2004}.  In prokaryotes, an analogous role is played by an important class of small non-coding RNAs (antisense sRNAs) that also act post-transcriptionally to negatively regulate proteins. {These antisense sRNAs can vary in size from tens to a few hundred nucleotides} \cite{Storz2011} and prevent translation by binding to the target mRNAs. 

While both {inhibitory} microRNAs and sRNAs play similar functional roles, they act by very different mechanisms \cite{BARTEL2004}.  Eukaryotic MicroRNAs undergo extensive post-processing and are eventually incorporated into the  RISC assembly \cite{Pillai2007,Kai2010}. The RISC complex containing the activated microRNA binds mRNAs and targets them for degradation. A single RISC complex molecule can degrade multiple mRNA molecules suggesting that microRNAs act catalytically to repress protein production. In contrast, both the mRNAs and sRNAs are degraded when bound to each other {in a process that can happen spontaneously} \cite{Waters2009,Song2008}{ or be mediated by extra machinery such as RNA chaperone Hfq}\cite{Lenz2004,Gottesman2011,Vogel2011}. {This suggests} that prokaryotic sRNAs, unlike their eukaryotic counterparts, act stoichiometrically on their targets.

Stoichiometric regulation has been extensively  studied theoretically and experimentally based on mass-action kinetics \cite{Levine2007, Levine2008, Mehta2008, Mitarai2009, Jia2009, Platini2011, Baker2012} {and experimental protocols have been proposed for determination of different modeling parameters}\cite{Elgart2010}. {On the other hand}, there exist relatively little {similar} work on understanding microRNAs and other catalytic non-coding RNAs \cite{Mukherji2011, Levine2007a, Hao2011}. Both theoretical calculations (see Supporting Information of  \cite{Levine2007}) and quantitative experiments  \cite{Mukherji2011} indicate that mean protein-expression of microRNA regulated genes follows a linear-threshold behavior similar to that of sRNAs in prokaryotes. {In contrast}, \cite{Hao2011}  {argued that stoichiometric sRNAs are noisier than catalytic microRNAs}. Presently, it is unclear how to reconcile these results and it raises the natural question of how the differing mechanisms employed by sRNAs and microRNAs affect the intrinsic noise profiles of regulated genes. 

To answer this question, we used a generalized model of gene regulation by {inhibitory} non-coding RNAs to calculate the mean expression and intrinsic noise of regulated proteins. Our model is based on stochastic mass-action kinetics with tunable parameters that allow us to vary the mode of action from  stoichiometric interactions such as those in prokaryotic sRNAs  to highly catalytic interactions that mimic eukaryotic microRNAs. {Our model is also applicable to plant microRNAs where microRNAs bind strongly to regulated mRNAs\cite{Jones-Rhoades2006}.} Contrary to \cite{Hao2011}, we show that in many parameter regimes the intrinsic noise properties of microRNAs and sRNAs are qualitatively similar. Finally, motivated by the competing endogenous RNA (ceRNA)  hypothesis which suggests that microRNAs induce an extensive mRNA-interaction network, we extended our model to consider the case where a microRNA regulates multiple mRNAs. We calculate the intrinsic noise for these models and show that noise is an extremely sensitive measure of crosstalk between mRNAs. This suggests a new experimental strategy for uncovering microRNA-induced mRNA interactions. {Recently we learned of three studies }\cite{Bosia2013, Figliuzzi2013 , Ala2013} {completed simultaneously with our work, in which they model multiple mRNAs interacting with multiple microRNAs. Bosia }{\it et al.} \cite{Bosia2013} {studied robustness in a noisy model of ceRNA, while Figliuzzi }{\it et al.}\cite{Figliuzzi2013} and Ala {\it et al.} \cite{Ala2013} {both focus on steady state behavior.  Figliuzzi} {\it et al.}\cite{Figliuzzi2013} {studied sensitivity of ceRNA to transcription rate changes using analytical techniques whereas Ala} {\it et al.} \cite{Ala2013} {combined analytical and bioinformatics results with experiments and showed cross-talk between ceRNAs. } Our work complements and extends these other works to analyze the noise characteristics of microRNA interactions. 
\sectionlarge{Results}

\begin{figure}[!ht]
\begin{center}
\includegraphics[width=0.9\textwidth]{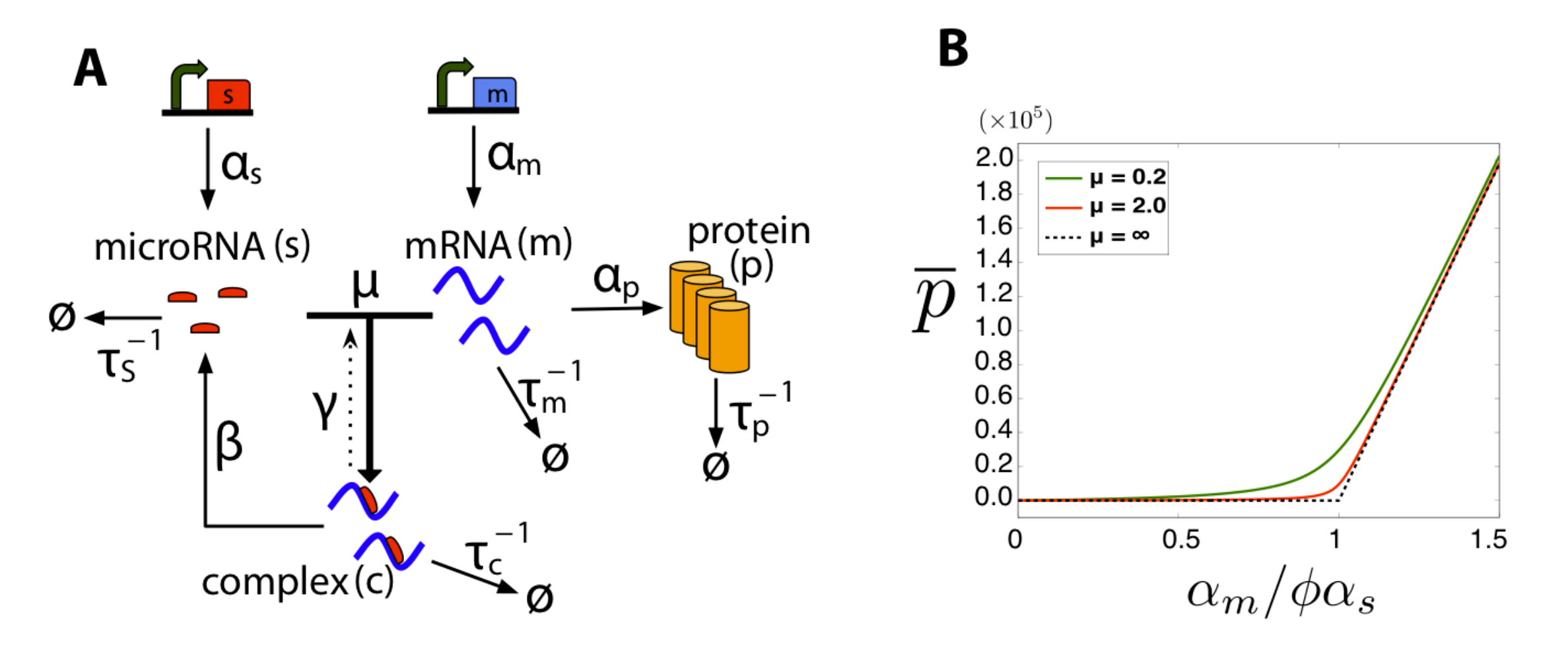}
\end{center}
\caption{\textbf{microRNA Model.} \textbf{A)} Schematic of the interactions between  noncoding RNAs and mRNA. $\alpha_s$ and $\alpha_m$ are          respectively the transcription rate of microRNA and mRNA. $\tau_m^{-1}, \tau_s^{-1}, \tau_c^{-1}, \tau_p^{-1}$ are respectively the degradation rates of mRNA,  noncoding RNA, complex and protein. $\mu$ and $\gamma$ are respectively the direct and reverse interaction rates between mRNA and  noncoding RNA, and $\beta$ is the catalyticity. \textbf{B)} Analytical results showing protein mean versus normalized transcription rate, $\frac{\alpha_m}{\phi\alpha_s}$, for different values of $\mu$ in the catalytic regime ($\beta=10, \tau_c=1$) where $\phi\equiv 1+\beta\tau_c$. The dashed line is the theoretical result for infinitely large $\mu$. The following parameters have been used in this plot: $\alpha_s=3, \alpha_p=4, \tau_s=30, \tau_m=10,  \tau_p=30, \gamma=1 $. {The stochastic simulations produce exactly the same mean (graph not shown here).}}
\label{fig:Intro}
\end{figure}

\subsectionmedium{Model Description}
Here we propose a model of gene regulation by non-coding RNAs based on mass-action kinetics \cite{Hao2011}. A schematic of the model is shown in Figure \ref{fig:Intro}A. Our model has four species: mRNA molecules denoted by $m$, functional non-coding RNAs denoted by $s$, mRNA-noncoding RNA complexes denoted by $c$, and the number of expressed proteins denoted by $p$.  We note that $s$ can represent either the {concentration} of prokaryotic small RNAs or the {concentration} of functional microRNAs found within the RISC complex. mRNA molecules are transcribed at the rates $\alpha_m$ and translated at a rate $\alpha_p$. {Free} mRNAs degrade at a rate $\tau_m^{-1}$. $\alpha_s$ represents the mean rate at which functional non-coding RNAs are formed. For prokaryotic sRNAs, this is simply the transcription rate of sRNAs. For microRNAs, this is an aggregate rate that accounts for the complicated kinetics of transcription and assembly into the RISC complex. mRNAs and noncoding RNAs form  complexes $c$ at a rate $\mu$ and disassociate at rate $\gamma$. Importantly, the complexes can also be degraded at a rate $\tau_c^{-1}$. This degradation can be actively regulated or conversely stem from dilution due to cell division. 

Once mRNAs bind noncoding RNA and form a complex $c$, they can no longer be translated, resulting in decreased protein expression. To account for the differences between microRNAs and sRNAs, we have an additional parameter $\beta$ which measures the ``recycling rate'' of noncoding RNA.  When  $\beta \gg \tau_c^{-1}, \gamma$, the non-coding RNAs function catalytically with a single non-coding RNA molecule able to bind and degrade multiple mRNA molecules\cite{Liu2004,Mallory2010}. This is a good model for gene regulation by microRNA in eukaryotes. In contrast, when $\beta  \ll \tau_c^{-1}$ the noncoding RNAs function stoichiometrically. In particular,  for prokaryotic sRNAs it is commonly believed that $\beta =0$ and no recycling of noncoding RNAs takes place. {In plants microRNA's sequence almost perfectly matches the mRNA sequence \cite{Jones-Rhoades2006}.  Plant microRNAs can be modeled using intermediate values of the recycling rate $\beta$ and  disassociation rate $\gamma$, although it is possible that they may behave stoichiometrically ($\beta\simeq0$) as is the case with bacterial small RNAs.} Finally, we note that the ratio of $\beta$ to $\gamma$ is a measure of how much of the gene regulation happens through nucleolytic cleavage in contrast to translational repression.
 
We use two different approaches to mathematically model gene regulation by non-coding RNAs. We calculate both the mean expression levels as well as ``intrinsic noise'' due to stochasticity in the underlying biochemical reactions.  First, we perform simple Monte-Carlo simulations for the reaction scheme outline above using the Gillespie algorithm\cite{Gillespie2007}. While  these simulations are exact, this method is computationally intensive making it difficult to systematically explore how noise properties depend on kinetic parameters.  For this reason, we use a second approach based on linear noise approximation (LNA)  \cite{Kampen2007, Swain2004, Mehta2008}. {The LNA approximates the exact master equations using Langevin equations. For very small particle numbers, this approximation breaks down even in determining the average particle numbers. Nonetheless, for medium and large particle numbers, the LNA is very adept at capturing qualitative behaviors of the system and we see a good agreement between LNA and exact simulations for our system for larger particle numbers and volumes. This allows us to derive analytical expressions for the noise and systematically explore how gene expression noise depends on system parameters.}

Within the LNA, we can mathematically represent our model using the equations
\begin{equation}
\begin{split}
\frac{ds}{dt}&=\alpha_{s}-\tau_{s}^{-1}s+\beta c+\gamma c-\mu ms+\eta_{s}+\eta_{\beta}+\eta_{\gamma}-\eta_{\mu}\\
\frac{dm}{dt}&=\alpha_{m}-\tau_{m}^{-1}m+\gamma c-\mu ms+\eta_{m}+\eta_{\gamma}-\eta_{\mu}\\
\frac{dc}{dt}&=\mu ms-\beta c-\gamma c-\tau_{c}^{-1}c+\eta_{c}-\eta_{\beta}-\eta_{\gamma}+\eta_{\mu}\\
\frac{dp}{dt}&=\alpha_{p}m-\tau_{p}^{-1}p+\eta_{p}
\end{split}
\label{eqn:SingleSpecies}
\end{equation}
{with $s, m, c, p$ being the concentration of noncoding RNA, mRNA, complex and protein respectively,} and $\eta_s, \eta_m, \eta_c$, and $\eta_p$, being the noise in the birth-death processes of non-coding RNAs, mRNAs, complex, and proteins respectively, $\eta_\mu$ the binding noise, $\eta_\gamma$ the unbinding noise, and $\eta_\beta$ the RNA recycling noise. The variance of these noise terms is given by
\begin{align}
\begin{split}
\overline{\eta_{i}(t)}&=0,\qquad i,j=s,\beta,\gamma,\mu,m,c,p\\
\overline{\eta_{i}(t)\eta_{j}(t+\tau)}&=\delta_{ij}\sigma_{i}^{2}\delta(\tau)
\end{split}
\end{align}
where $\delta_{ij}$ is the Kronecker delta {and overline represents time-average. For the noise terms we have:}
\begin{align}
\sigma_{s}^{2}=\alpha_{s}+\tau_{s}^{-1}\overline{s},\quad\sigma_{\beta}^{2}=\beta \overline{c},\quad\sigma_{\gamma}^{2}=\gamma \overline{c},\quad\sigma_{\mu}^{2}=\mu \overline{m} \,\overline{s},\\\nonumber
\sigma_{m}^{2}=\alpha_{m}+\tau_{m}^{-1}\overline{m},\quad\sigma_{c}^{2}=\tau_{c}^{-1}\overline{c},\quad\sigma_{p}^{2}=\alpha_{p}\overline{m}+\tau_{p}^{-1}\overline{p},
\end{align}
with $\overline{n}$ the {time-averaged} steady-state concentration of species $n$. {Note we can convert between particle number and concentration by multiplying by appropriately chosen volumes (see Material and Methods). } We modeled each interaction in Figure \ref{fig:Intro}A  as an independent Poisson process with a mean rate given by  mass-action kinetics\cite{Kampen2007}. Fluctuations in each interaction have been modeled by an independent Langevin term. Care must be taken to ensure the correct sign in the cross-correlations \cite{Swain2004}. In the remainder of the paper, analytical results from the linear noise approximation are shown along with simulations using the Gillespie algorithm \cite{Gillespie2007}. Both methods are in good agreement.

{Finally, in what follows we will focus on protein mean and noise. However note that protein in our model is dependent on mRNA through a simple linear birth-death process. In this sense protein can be thought of as a readout of mRNA that goes through a low-pass filter set by $\tau_p^{-1}$. As a result the general behavior of protein and mRNA is very similar in our model and lead to similar qualitative results.}

\subsectionmedium{Mean Expression Levels}
We begin by deriving the steady-state response of our system by setting the time-derivatives of the left hand side of equation \eqref{eqn:SingleSpecies}  to zero. Denoting the steady-state concentration of a species $n$ by $\overline{n}$, we find that
\begin{align}
\begin{split}
& \overline{s}=\frac{\phi\alpha_{s}-\alpha_{m}-\lambda+\sqrt{\left(\phi\alpha_{s}-\alpha_{m}-\lambda\right)^{2}+4\lambda\phi\alpha_{s}}}{2\phi\tau_{s}^{-1}}\\
 & \overline{m}=\frac{\alpha_{m}-\phi\alpha_{s}-\lambda+\sqrt{\left(\alpha_{m}-\phi\alpha_{s}-\lambda\right)^{2}+4\lambda\alpha_{m}}}{2\tau_{m}^{-1}}\\
&\overline{c}=\mu \overline{m} \,\overline{s}\tau_{cR}\\
&\overline{p}=\alpha_p\tau_p \overline{m}
\end{split}
\label{eqn:meanvalues}
\end{align}
 where we have defined the quantities
\be
\phi=1+\beta\tau_c,
\ee
\be
\tau_{cR}^{-1} = \beta + \gamma + \tau_c^{-1}
\ee 
\be
q = \frac {\tau_c^{-1}}{\tau_{cR}^{-1}} = \frac {\tau_c^{-1}}{\beta + \gamma + \tau_c^{-1}}
\label{defq}
\ee
\be
\lambda=\frac{1}{\mu\tau_{m}\tau_{s}q},\quad 
\label{deflambda}
\ee

{Note that to derive equation} \eqref{eqn:meanvalues}{ we have set the noise terms in equation }\eqref{eqn:SingleSpecies}{ to zero. This effectively eliminates the contribution of cross-correlation terms into steady-state solutions (i.e. $\overline{ms}$ is approximated as $\overline{m} \, \overline{s}$). For very small particle numbers, where LNA breaks down, this approximation also breaks down. This can especially happen for small system volumes. However we will not study such parameter regimes in this paper  }(see {\it Materials and Methods}). Furthermore, {we confine ourselves to the biologically-relevant parameter sets where the mRNA-microRNA association rate, $\mu$, is much larger than degradation rates of RNAs ($\tau_m^{-1},\tau_s^{-1}$) and lambda is small ($\lambda\ll 1$). For these parameters, there is a clear linear-threshold behavior in mean protein concentration (Figure 1B).  Such behavior has been extensively studied in the context of non-coding RNAs where gene expression is often divided into three distinct regimes: a repressed regime, an expressing regime, and a crossover regime } \cite{Levine2007, Mehta2008, Mukherji2011}. In the repressed regime ($\alpha_m\ll \phi \alpha_s$), non-coding RNAs almost always bind mRNAs and prevent translation.  In contrast, in the expressing regime ($\alpha_m\gg \phi \alpha_s$) there are many more mRNAs than noncoding-RNAs, resulting in protein production that varies linearly with $\alpha_m$. Finally, there is a crossover between these two behaviors when $\alpha_m\simeq \phi\alpha_s$. 

The factor $\phi$ multiplying $\alpha_s$ renormalizes the transcription rate of the non-coding RNA to account for the fact that {a} single non-coding RNA can degrade multiple mRNAs. To see this note that $\phi^{-1}= \tau_c^{-1}/(\beta+\tau_c^{-1})$ is just the probability that a non-coding RNA is degraded when it is bound to an mRNA in a complex under the assumption complexes do not disassociate ($\gamma\ll\beta$ or $\gamma \ll  \tau_c^{-1}$). Thus, we can think of $\phi$ as the average number of mRNAs that a non-coding RNA will degrade before it is itself degraded. As expected, when $\beta=0$, $\phi=1$ and these results reduce to those derived for prokaryotic sRNAs \cite{Levine2007, Mehta2008}.  {Overall the experimentally observed linear threshold behavior of mean protein in quantitative single-cell experiments }\cite{Mukherji2011} {is consistent with our theoretical calculation of the mean.}

\begin{figure}[!ht]
\begin{center}
\includegraphics[width=0.9\textwidth]{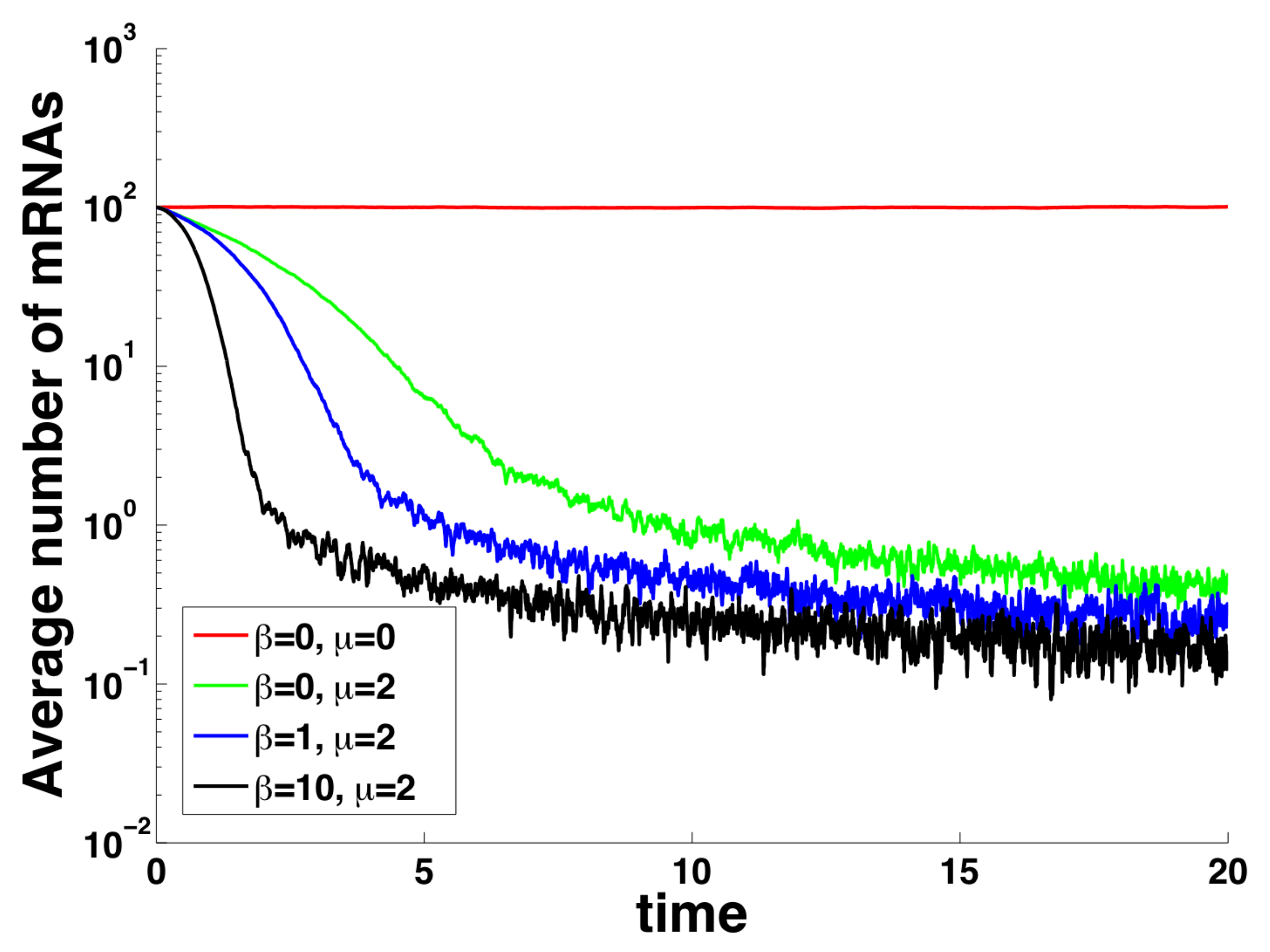}
\end{center}
\caption{{\textbf{mRNA Decay.}} Time-course of mean mRNA number showing different decay rates in different parameter regimes. Each curve is the average of 100 Gillespie simulations with the same initial conditions starting at {steady state of the unregulated model ($m=\alpha_m\tau_m, s=0, c=0$)}. Parameters same as Figure \ref{fig:Intro} with $\alpha_m=1, \beta=10$.}
\label{fig:TimeCourse}
\end{figure}

{Comparing decay patterns of mRNAs repressed by microRNAs to free mRNAs provides a method to determine the existence and efficiency of microRNA binding. For example, Braun} {\it et al.} \cite{Braun2011} {use immunoblotting to compare eukaryotic cells with no microRNA to cells transfected with microRNA transcripts and show a two-fold decrease in mRNA lifetime as a result of microRNA interaction. To determine if the general decay pattern of mRNAs in our model mimics their results{, we studied the effect of microRNAs in the transient behavior of mRNAs.} Figure }\ref{fig:TimeCourse} {shows the results. To mimic the population averaging inherent to immunoblots, each curve is calculated by averaging 100 Gillespie simulations starting from the same initial conditions. Note that once the microRNA is introduced into the system (by setting $\mu=2$) the lifetime of mRNA dramatically decreases. Changing $\beta$ from 0 (stoichiometric) to 10 (catalytic) further decreases the lifetime. The fluctuations in particle number close to zero are due to the finite ensemble size.{ In this regard our results qualitatively agree with experiments in }\cite{Braun2011} {regardless of the regulatory regime studied. Furthermore we observe a fast initial decay rate followed by slower decay at later times instead of a single exponential decay. This is in agreement with the observation made in }\cite{Braun2011}.{ However we believe that beyond such qualitative agreement, a precise fit to the experimental data would require more complicated models. It is well known that the degradation of mRNAs includes several stages. In a recent work by Deneke et al.} \cite{Deneke2013}{ it was shown that inclusion of multistep degradation can lead to piece-wise exponential decay of mRNAs. We speculate that such a multistep process in conjunction with the mechanism proposed here can lead to the correct quantitative decay patterns observed experimentally.}}
\subsectionmedium{Intrinsic Noise}

\begin{figure}[!ht]
\begin{center}
\includegraphics[width=0.9\textwidth]{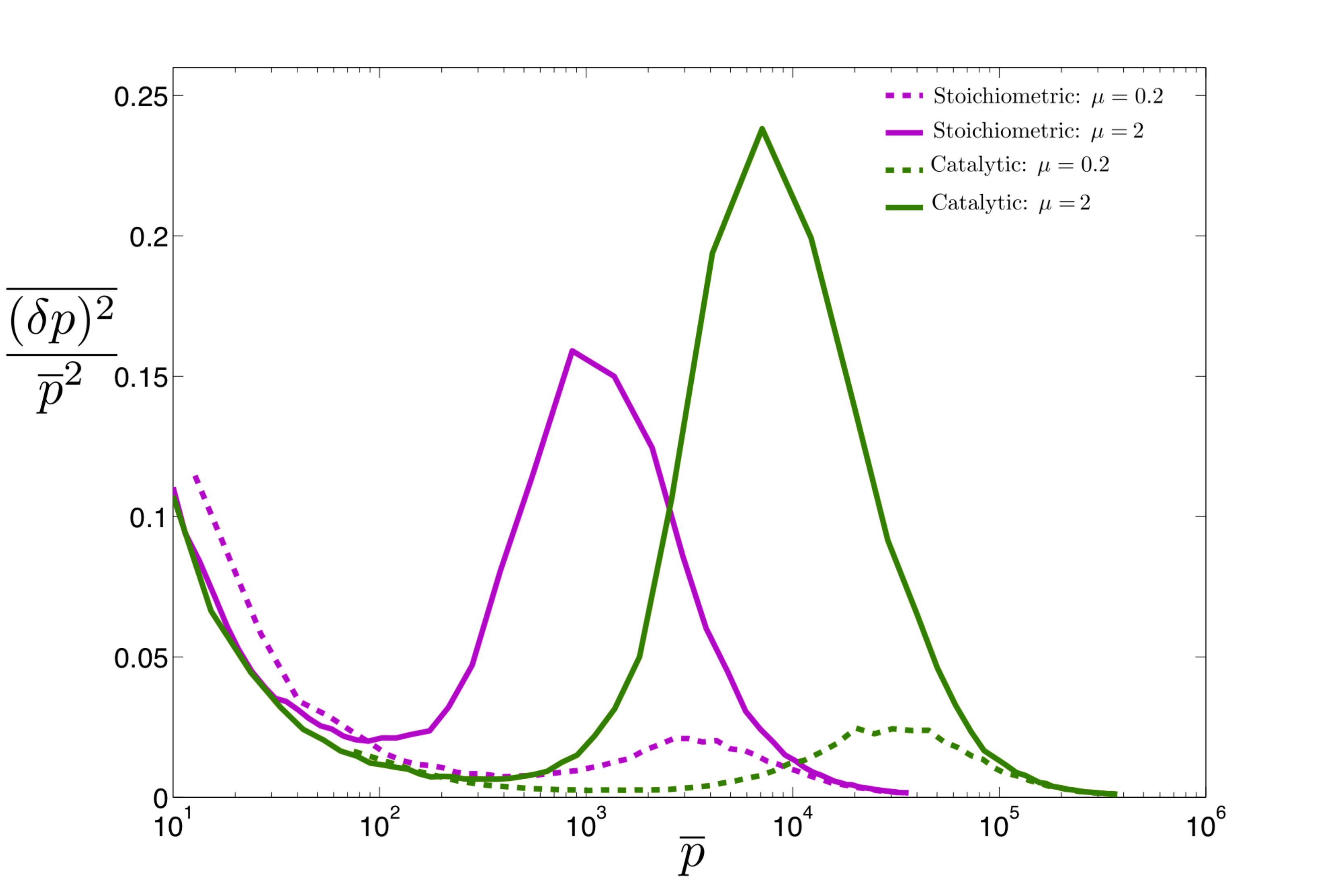}
\end{center}
\caption{\textbf{Protein Noise in Two Regimes.} {Gillespie simulation} results showing protein noise as a function of mean protein concentration for stoichiometric and catalytic interactions {plotted for two different values of interaction rate $\mu$}. For {each value of $\mu$} {catalytic interaction has a slightly higher noise in the crossover regime compared to stoichiometric interaction, otherwise a similar three-regime behavior can be observed in both cases. We have plotted this result for two different values of $\mu$ to show that this observation is qualitatively independent of interaction strength, and only affects the level of noise.}  Parameters same as in Figure \ref{fig:Intro}B. Stoichiometric regime with $\beta=0$ and catalytic regime with $\beta=10$. }
\label{fig:AllRegimes}
\end{figure}

Gene regulation is inherently stochastic due to the small number of molecules present in the cell. Noise has important phenotypic consequences for a variety of biological phenomena \cite{Balazsi2011} and for this reason it is important to characterize the intrinsic noise properties of non-coding RNAs.  As is usual, we define the intrinsic noise as the variance in protein level divided by mean protein level squared, $\frac{\overline{\left(p-\overline{p}\right)^2}}{\overline{p}^2}$, where overline represents steady state time averaging. This is a measure of relative protein fluctuations compared to their mean. Study of noise in bacterial sRNAs shows a peak in the crossover regime that emerges as a result of competition between mRNA and sRNA \cite{Mehta2008}. The switch-like behavior of the system, due to its linear-threshold nature, results in an amplification of any fluctuations in the mRNA level that is in excess of the mRNAs bound to noncoding-RNA. As a result, noise is increased in the crossover regime. 
We observe a similar qualitative behavior for noise in all parameter regimes of our model {obtained by computing the solution to  equation }\eqref{eqn:SingleSpecies} {at steady state}. Figure \ref{fig:AllRegimes} is a plot of protein noise as a function of mean protein concentration for catalytic and stoichiometric interactions, showing a peak at protein levels that correspond to the crossover regime in both cases. The height of the peak increases with the binding affinity $\mu$ of a non-coding RNA for its target mRNA. On a whole, the noise profile of catalytic and non-catalytic genes is remarkably similar. There are however slight differences. The peak is slightly higher and occurs at a slightly larger mean protein level for catalytic non-coding RNAs. This suggests that the underlying reason for the noise peak is not the enzymatic mode of action of non-coding RNAs, but the fact that the number of mRNAs and  the number of non-coding RNAs (appropriately normalized by $\phi$) are almost equal.

\begin{figure}[!ht]
\begin{center}
\includegraphics[width=0.9\textwidth]{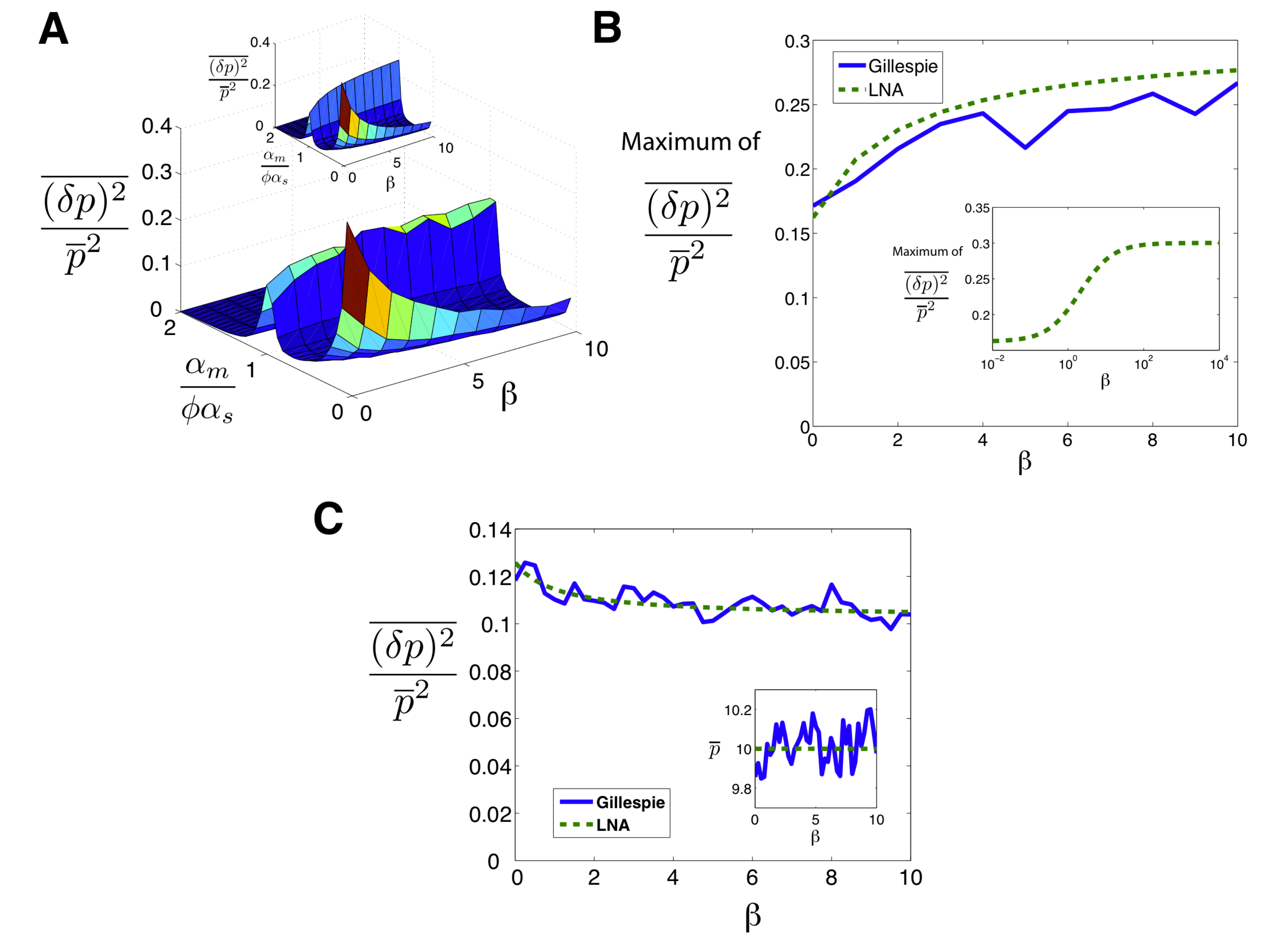}
\end{center}
\caption{\textbf{Effect of Catalyticity on Noise.}  \textbf{A)} Simulation results for noise as a function of $\beta$ and $\frac{\alpha_m}{\phi\alpha_s}$. Inset: analytical results for the same system. Parameters same as Figure \ref{fig:Intro} with $\mu=2$. {\textbf{B)} Maximum of noise in the crossover regime as a function of $\beta$. Inset: Same result shown using analytical method for larger range of parameters showing the plateau (unaccessable computationally due to large particle numbers). \textbf{C)} Noise in the repressed regime as a function of $\beta$ for constant protein mean ($\overline{p}=10$). Inset: protein mean as a function of $\beta$. }}
\label{fig:ChangingBeta}
\end{figure}

To better understand the effect of catalytic interaction on noise, we calculated protein noise versus $\alpha_m$ for different values of $\beta$. Figure \ref{fig:ChangingBeta}A shows the results for Gillespie simulation and linear noise approximation {(the inset) as a function of $\alpha_m\over\phi\alpha_s$ and $\beta$ showing that the three-regime noise behavior is robust over a large range of $\beta$. Note that in all cases,  the noise is peaked in the crossover regime. Aside from small discrepancies caused by the approximate nature of the analytical method as well as the finite statistical sample used in simulations,  we see a good qualitative agreement between the two methods.   To better visualize the robustness of noise to changes in $\beta$ we have plotted the maximum of noise (at crossover regime) as a function of $\beta$ in Figure }\ref{fig:ChangingBeta}B{. Notice that the peak height initially increases as a function of $\beta$ and reaches a plateau for large $\beta$ upon entering the catalytic regime. However there are not any significant differences in the two interaction regimes. Finally as another test of the robustness of the model to parameter changes, we studied noise of a catalytic (i.e. eukaryotic microRNA, $\beta\gg\tau_c^{-1}, \gamma$) and stoichiometric interaction (i.e. prokaryotic sRNA, $\beta=0$) for the same level of mean protein produced.  Figure} \ref{fig:ChangingBeta}C {shows this comparison for noise in the repressed regime as a function of $\beta$ with $\alpha_m$ chosen such as to keep mean protein concentration constant. As can be seen, the noise decreases from its original value in the stoichiometric regime ($\beta=0$) to a slightly lower value as $\beta$ is slightly increased ($\beta\simeq 2$) and any further increase in $\beta$ does not affect noise dramatically. This again reveals that the system is robust to changes in $\beta$.}

\begin{figure}[!ht]
\begin{center}
\includegraphics[width=1\textwidth]{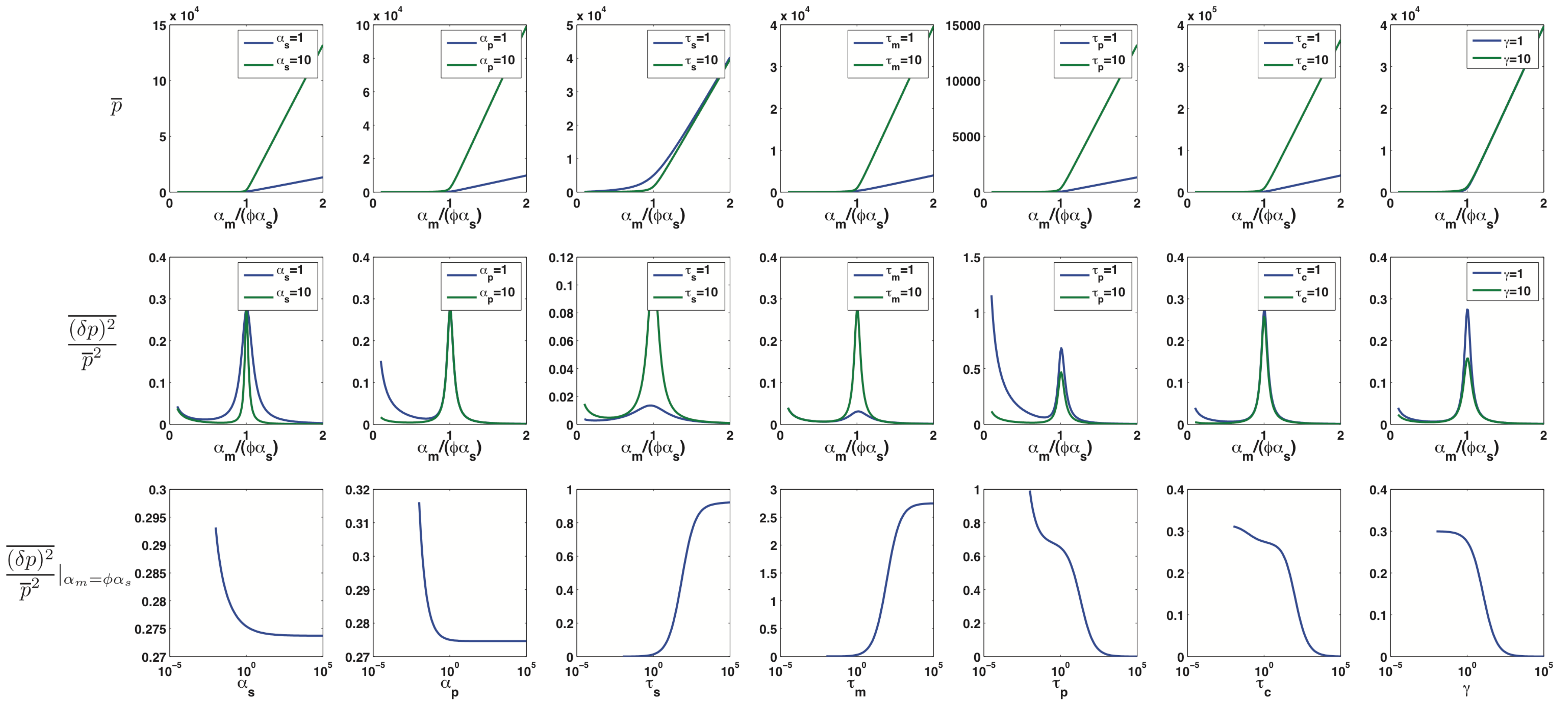}
\end{center}
\caption{{\textbf{Qualitative Robustness of Mean and Noise.}}  First and second row show analytical results for protein mean and noise respectively as a function of $\alpha_m\over \phi\alpha_s$ while in each column one single parameter is varied. Third row is analytical results for protein noise calculated at $\alpha_m=\phi\alpha_s$ and plotted as a function of the parameter under study. Parameters that are not changed in each graph are the same as in Figure \ref{fig:Intro}.}
\label{fig:ParameterChanges}
\end{figure}

{ As a final test of the robustness of system on parameters, we studied protein mean and noise for a range of parameters using LNA. Figure} \ref{fig:ParameterChanges}  {shows the results with first and second row corresponding to protein mean and noise respectively. Note that the linear-threshold behavior of mean and the peak at crossover regime is consistent for a wide range of parameters, showing that these qualitative features are independent of our choice of parameters, given that the number of particles are large enough for LNA to hold. The third row in figure }\ref{fig:ParameterChanges} {shows noise at the crossover regime ($\alpha_m=\phi\alpha_s$) plotted for a wide range of paramters. We see a monotonic change in noise as different parameters are varied showing that changes in parameters can change the exact  quantitative results but that the qualitative behavior of the intrinsic noise is similar for almost all biologically realistic choices of parameters. }

\subsectionmedium{Including Transcriptional Bursting}
Experimental  evidence suggests  that  mRNAs are often produced in bursts \cite{Golding2005}. Transcriptional bursting represents another important source of stochasticity that was ignored in the analysis presented above. We can extend the model presented above to  incorporate transcriptional bursting by  considering the case where the genes encoding for mRNAs can be in two distinct states: an ``on'' state  where mRNAs can be transcribed and an ``off'' state where transcription is not possible.  For example, in eukaryotes the two states may correspond to whether the chromatin is condensed or not \cite{Kaern2005}. To model transcriptional bursting we replace the equation for mRNA production in Eq. \eqref{eqn:SingleSpecies} by the pair of equations
\begin{align}
\frac{dg}{dt}&=k_-(1-g)-k_+g+\eta_g\\\nonumber
\frac{dm}{dt}&=\alpha_m^{on}g-\tau_{m}^{-1}m+\gamma c-\mu ms+\eta_{m}+\eta_{\gamma}-\eta_{\mu}
\end{align}
where the probability that a gene is in the on state is denoted by $g$, $k_-$($k_+$) are the on(off) rate of the gene, and $\alpha_{m}^{on}$ is the maximum mRNA transcription rate.  {The average state of the gene} is given by $\overline{g}= k_-/(k_- + k_+)$\cite{Mehta2008}. The gene noise $\eta_g$ is Gaussian white noise with mean zero and variance given by $\overline{\eta_{g}(t)\eta_{g}(t+\tau)}=2k_+\overline{g}\delta(\tau)$. The mRNA noise $\eta_m$ is now $\overline{\eta_{m}(t)\eta_{m}(t+\tau)}=(\alpha_{m}^{on}\overline{g}+\tau_{m}^{-1}\overline{m})\delta(\tau)$. All other equations remain the same as before.

\begin{figure}[!ht]
\begin{center}
\includegraphics[width=0.9\textwidth]{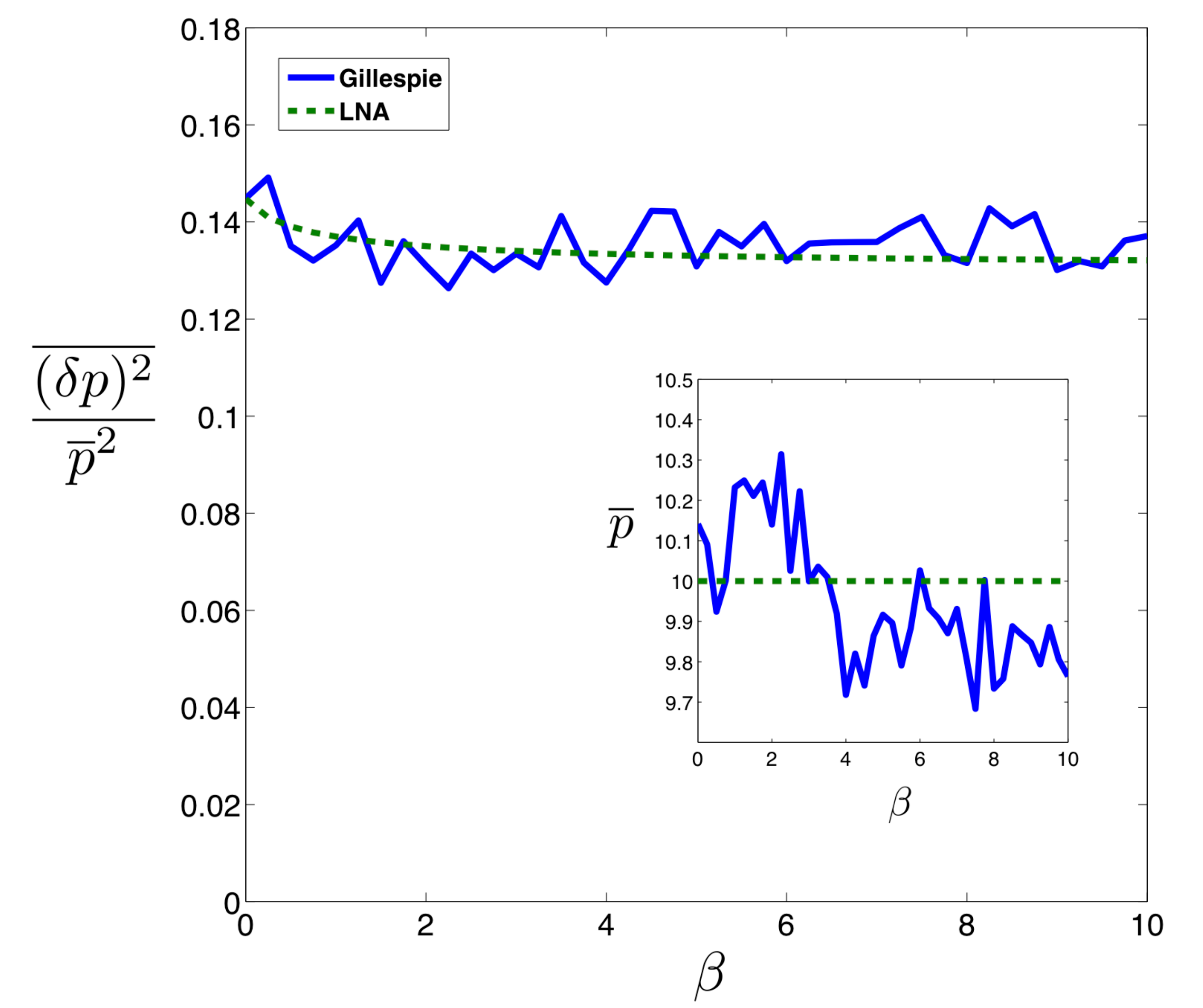}
\end{center}
\caption{\textbf{Catalyticity and Bursting.} Noise in the repressed regime with bursting as a function of $\beta$ for constant protein mean. For each data point $\alpha_m$ is chosen such that $<p>=10$. Furthermore $\alpha_m^{on}=10, k_{-}=1, k_{+}=k_-(\frac{\alpha_m^{on}}{\alpha_m}-1)$. The remaining  parameters are same as in Figure \ref{fig:Intro} with $\mu=2$. Inset: protein mean as a function of $\beta$.}
\label{fig:RepressedBursting}
\end{figure}

In the analysis that follows,  we assume that $k_-$ is fixed but $k_+$ can vary. This corresponds to the biological situation where a gene is regulated by a repressor that can turn the gene off. To compare noise for different values of $\beta$,  we choose $k_+$ so that the mean protein levels remain constant. Furthermore, we concentrate on the repressed regime (see Figure \ref{fig:RepressedBursting}). Here noise decreases slightly as $\beta$ is increased which is very similar to the case without bursting as was shown in Figure \ref{fig:ChangingBeta}B. This means that noise in the repressed regime is insensitive to bursting regardless of how catalytic the interaction is. {This is true as long as the microRNA-mRNA binding rate is large compared to RNA lifetimes and mean protein levels exhibit the biologically observed linear-threshold behavior.}

{However note that it is possible to design a very slow gene with the same mean value as here, by choosing small on and off rates, $k_-/k_+$, while keeping their ratio constant. In such cases, if the dynamics of the gene are chosen much slower than the dynamics of the rest of the network, effectively the system will switch between two different transcription rates corresponding to two different linear threshold behaviors. As a result such a system would behave as the superposition of two networks with their crossover regimes happening at different transcription rates, leading to a bimodal distribution of protein noise. It has been shown experimentally that in some biological systems transcription happens in this regime. For example Raj }{\it et al.}\cite{Raj2006} {studied transcription through bursting in mammalian cells and discovered a bimodal protein distribution. Using stochastic modeling }\cite{Peccoud1995}, { they showed that their results agree with the notion of a slower gene activation process. In this work we have limited ourselves to faster gene dynamics where distributions are unimodal. However, as a future direction, it would be interesting to study the effect of noncoding RNAs on the time-scales of the process in different parameter regimes. If these time-scales remain small compared to gene activation/deactivation time-scale, protein noise will exhibit a bimodal distribution. We speculate that in this case instead of the three-regime behavior discussed here, the system will exhibit five (or four) regulatory regimes depending on the existence of (or lack of) leaky genes.}
 
\subsectionmedium{Asymptotic Formulas for Noise}
\subsubsectionsmall{Expressed}
To gain further insight, we have derived asymptotic formulas for the noise in the repressed and expressing regime. In the expressing regime with large mRNA transcription, we define the small parameter $\epsilon \equiv\frac{\lambda}{\alpha_m-\phi\alpha_s}$ with $\lambda$ given by Eq. \ref{deflambda}.  In the expressing regime, $\epsilon \ll 1$, and the steady-state expression levels of mRNA and non-coding RNAs take the form
\begin{equation}
\overline{m}\simeq \frac{1}{q\mu\tau_s}\frac{1}{\epsilon},\quad\overline{s} \simeq\alpha_s\tau_s\epsilon\\
\end{equation}
with $q$ given by Eq. \ref{defq}. Furthermore, the protein  noise in this regime is identical to that of a transcriptionally regulated  gene {without post-transcriptional regulation}\cite{Thattai2001} and is given by 
\begin{align}
\frac{\overline{(\delta p)^2}}{\overline{p}^2}\simeq  {1+b\frac{\tau_p}{\tau_p+\tau_m} \over \overline{p} } \simeq {1+b \over {\overline{p}}}
\end{align}
where {$\delta p\equiv p-\overline{p}$ and } $b\equiv\alpha_p\tau_m$ (see {\it Materials and Methods}) and it is assumed that mRNAs degrade much faster than proteins ($\tau_p\gg \tau_m$). {In simple models of gene expression  where mRNAs are produced stochastically with a Poisson rate $\alpha_m$, the quantity $b$ is often called the `burst size'  and measures the average number of proteins made from a single mRNA molecule} \cite{Swain2002}. {Our result suggest that in the expressing regime protein noise is independent of $\beta$  and the noise profile of  post-transcriptionally regulated genes is similar to those of genes regulated via transcription factors.} \cite{Thattai2001}. 

\subsubsectionsmall{Repressed}
In the repressed regime, we now have  $\epsilon \equiv\frac{\lambda}{\phi\alpha_s-\alpha_m}  \ll 1$ and the average number of mRNA and non-coding RNA molecules is given by
\begin{align}
\overline{m}&\simeq\alpha_m\tau_m\epsilon\\\nonumber\overline{s}&\simeq\frac{1}{q\mu\tau_m\phi\epsilon}
\end{align}
where $q$ 
\begin{align}
q\equiv\frac{\tau_{c}^{-1} }{\beta+\gamma+\tau_{c}^{-1}},
\end{align}
is the probability that a complex is degraded.
The noise in this regime is given by
\be
\frac{\overline{(\delta p)^2}}{{\overline{p}}^2}\simeq \frac{1+\zeta b\epsilon}{\overline{p}}
\ee
where   $\zeta$ is a constant that is dependent on both $\beta$ and $\gamma$ (see {\it Materials and Methods}). When $\beta$ or $\tau_c^{-1} \gg \gamma$ it can be shown that $\zeta \simeq 1$. Note that this condition includes the catalytic regime with  $\beta\gg \gamma, \tau_c^{-1}$ as well as  the stoichiometric regime $\beta=0, \tau_c^{-1} \gg \gamma$ (see  {\it Materials and Methods}).    Thus, we conclude that in the repressed regime, non-coding RNAs reduce noise.  In particular, proteins are now produced from mRNAs in a smaller burst with typical size given by  $ \epsilon \zeta b\ll b$. Since the burst size is just the average number of proteins made from an mRNA, $b=\alpha_b \tau_m$, we can equivalently interpret this results as changing the effective lifetime of the mRNA molecules from $\tau_m$ to $\epsilon \zeta \tau_m$ \cite{Levine2007 , Mehta2008}. {This interpretation is consistent with the stochastic simulations on mRNA lifetimes  shown in Figure} \ref{fig:TimeCourse}.

\begin{figure}[!ht]
\begin{center}
\includegraphics[width=0.9\textwidth]{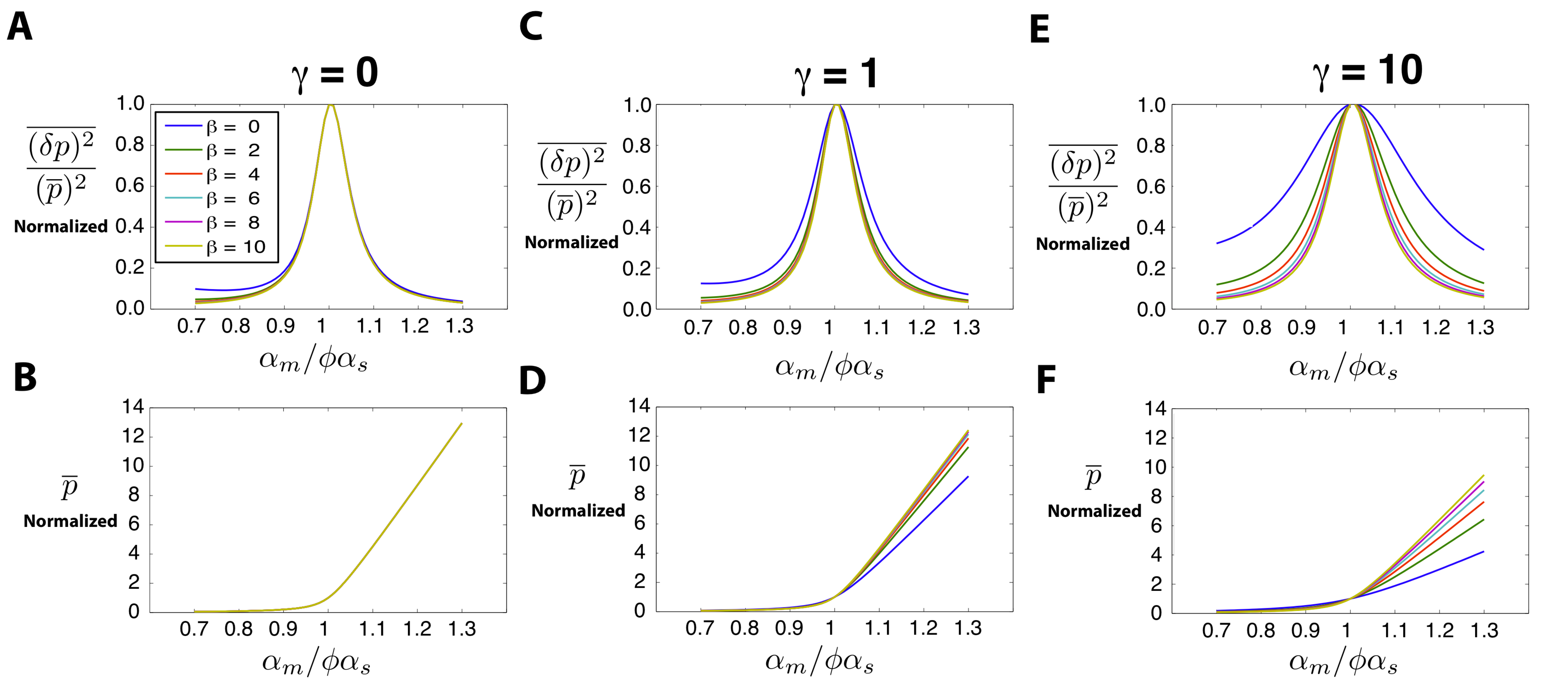}
\end{center}
\caption{\textbf{Scaling at Crossover Regime.} Parameters same as Figure \ref{fig:Intro} with $\mu=2$. {\bf A)} Protein noise normalized to its value at $\alpha_m=\phi\alpha_s$ {(more explicitly ${\overline{(\delta p)^2}\over(\overline {p})^2} / \left[{\overline{(\delta p)^2}\over(\overline {p})^2}|_{{\alpha_m\over \phi\alpha_s}=1}\right]$)} plotted as a function of $\alpha_m\over\phi\alpha_s$ for $\gamma=0$. Each line is a different value of $\beta$. Same legend for all figures. {\bf B)} Protein mean normalized to its value at $\alpha_m=\phi\alpha_s$ {(more explicitly $\overline {p}/ \left[\overline {p}|_{{\alpha_m\over \phi\alpha_s}=1}\right]$)}plotted as a function of $\alpha_m\over\phi\alpha_s$ for $\gamma=0$. {\bf C,D)} Graphs similar to  {\bf A,B} plotted for $\gamma=1$.  {\bf E,F)} Graphs similar to  {\bf A,B} plotted for $\gamma=10$.}
\label{fig:Scaling}
\end{figure}

\subsectionmedium{Scaling Behavior Near Crossover Regime}
Our analysis show that the width of noise peak at the crossover regime is independent of recycling ratio $\beta$. To understand this behavior better, we studied the crossover regime for different parameter values using linear noise approximation {(for comparison to simulations see {\it Materials and Methods})}. Figure \ref{fig:Scaling} shows the protein noise and protein mean  as a function of $\alpha_m\over \phi\alpha_s$ after rescaling both by their value at the crossover ($\alpha_m = \phi\alpha_s$)  for various recycling ratios. Notice that for $\gamma\ll\tau_c^{-1}$ these normalized plots of protein mean and noise show an approximate data collapse. As $\gamma$ is increased, this scaling behavior breaks down (Figure \ref{fig:Scaling}). The collapse of data for mean protein can be analytically derived given the fact that mean protein is only dependent on $\beta, \gamma, \tau_c$ through the combination $q\phi\equiv {\beta+\tau_c^{-1} \over \tau_c^{-1}+\beta+\gamma}$. 

This scaling of the mean protein number can be better understood if we define $x\equiv{\alpha_m\over \phi\alpha_s}$ and define {$\overline{p(x)}$} as the mean number of proteins corresponding to this value. Dividing this quantity by its value at $x=1$ {(mean protein at crossover)} and using equation \eqref{eqn:meanvalues} results in the following equation for the normalized mean protein:
\begin{align}
{\quad\overline{p(x)}\quad\over \overline{p(1)}}=\frac{x-1-{\nu\over q\phi}
+\sqrt{\left(x-1-
{\nu\over q\phi}\right)^{2}+4{\nu\over q\phi}{x}}}{{-{\nu\over q\phi}
+\sqrt{\left(
{\nu\over q\phi}\right)^{2}+4{\nu\over q\phi}{}}}}.\\\nonumber
\end{align}
where $\nu\equiv{\tau_m^{-1}\tau_s^{-1}\over\mu\alpha_s}$ is a constant with no dependence on $\beta, \gamma$ or $\tau_c$. Thus, the normalized mean protein number depends on $\beta, \gamma, \tau_c$ only through the combination $q\phi\equiv {\beta+\tau_c^{-1} \over \tau_c^{-1}+\beta+\gamma}$. The parameter $q \phi$ is the probability a complex will disassociate. In the limit $\gamma\ll\beta,\tau_c^{-1}$ this parameter will be equal to 1 and the scaled mean protein level becomes independent of $\beta$ causing the curves for different $\beta$ to collapse on top of each other (Figure \ref{fig:Scaling}). It is also interesting to note that any other condition on the parameters that removes the dependence of $q\phi$ on $\beta$ will also have the same effect (e.g. $\gamma, \beta\ll\tau_c^{-1}$). Somewhat more surprisingly, {near crossover} the noise also shows an approximate data collapse. We suspect that this collapse is likely indicative of universality near the crossover between the repressed and expressing regimes.

\subsectionmedium{mRNA Crosstalk and the ceRNA Hypothesis} 

Recently, the competing endogenous RNA (ceRNA) hypothesis proposed that microRNAs may play a crucial role in the cell in global gene regulation by inducing interactions between mRNA species \cite{Salmena2011}. The central mechanism underlying the ceRNA hypothesis is the idea that mRNA species may have interactions amongst themselves that are not direct but are instead indirect and mediated by competition  and depletion of shared microRNA pools.  The hypothesis is that these indirect mRNA interactions {result} in a biologically important mRNA network. However, the breadth and strength of microRNA induced interactions in eukaryotic genomes is still not well understood. For this reason, it is crucial to develop new strategies for measuring microRNA induced interactions between commonly regulated mRNAs. To this end, we asked whether the noise profile of regulated mRNAs could be used to uncover microRNA induced interactions. As a first step, we studied the simplified case where two different mRNA species are regulated by a single microRNA and  compete over a common pool of microRNAs, and we focused on the effect of  microRNA-induced crosstalk between mRNA species on the noise properties of regulated genes. The results presented here can be easily generalized to the case of many mRNAs interacting with many microRNAs.

\begin{figure}[!ht]
\begin{center}
\includegraphics[width=0.9\textwidth]{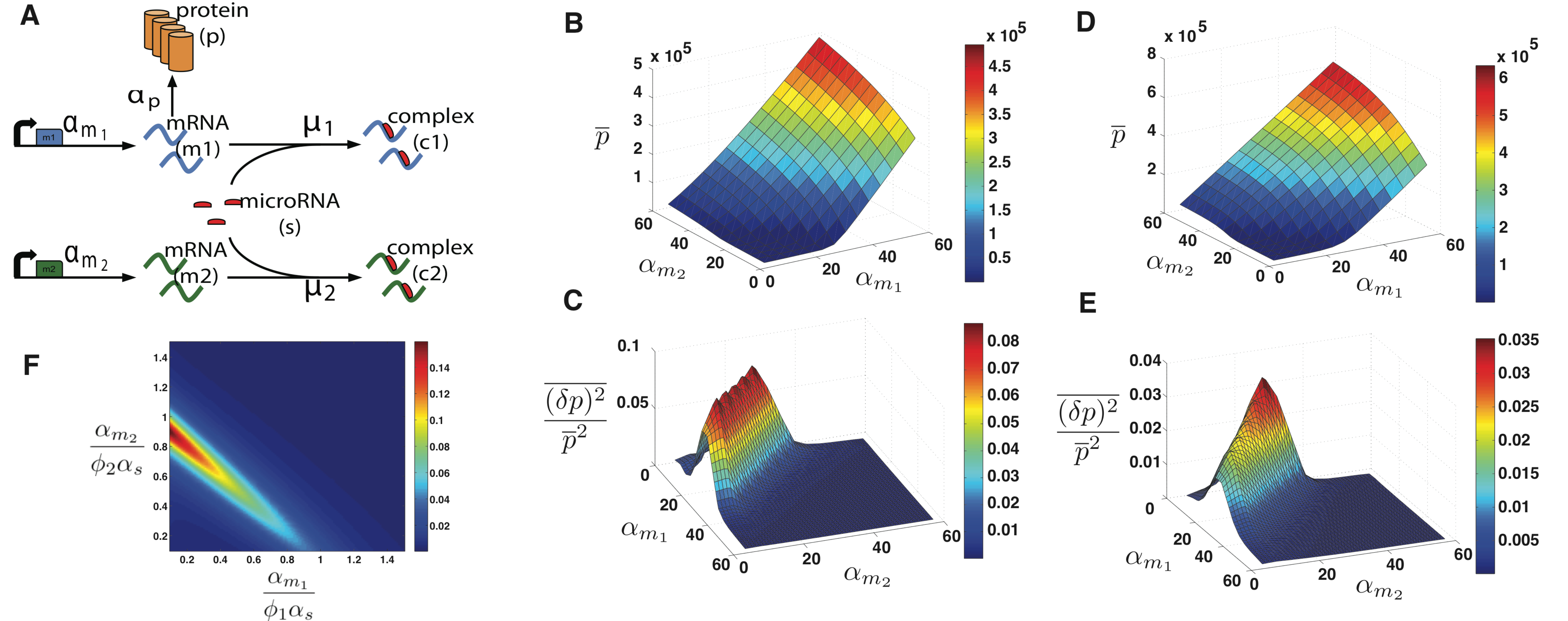}
\end{center}
\caption{\textbf{ceRNA Hypothesis.}  \textbf{A)} Schematic of mRNA crosstalk through a shared pool of microRNAs. $\alpha_{m_{1,2}}$ stand for transcription rate of each mRNA and $\mu_{1,2}$ correspond to interaction rates between mRNA and microRNA. The other interactions are as in Figure 1A. \textbf{B,C)} {Gillespie results for } protein mean (B) and noise (C) as a function of transcription rates of the two mRNAs with equal $\mu's$. Parameters same as in Figure \ref{fig:Intro} with $\mu_1=\mu_2=2$. \textbf{D,E)} {Gillespie results for } protein mean (D) and noise (E) as a function of transcription rates of the two mRNAs with unequal $\mu's$. Parameters same as in Figure \ref{fig:Intro} with $\mu_1=0.2, \mu_2=2$. Noise surface plots have been smoothed and interpolated for better visibillity. {\textbf{F)} LNA results for noise of two non-identical species as a function of normalized transcription rates (with $\phi_i\equiv 1+\beta_i\tau_{c_i}$)}. All the parameters are chosen to be distinct, i.e. $\alpha_s=3, \alpha_{p_1}=4, \alpha_{p_2}=10, \tau_s=30, \tau_{m_1}=10, \tau_{m_2}=20, \tau_{c_1}=1, \tau_{c_2}=5, \tau_{p_1}=30,  \beta_1=10, \beta_2=1, \gamma_1=1, \gamma_2=10, \mu_1=0.2, \mu_2=2, V=10$.}
\label{fig:ceRNA}
\end{figure}

A schematic of our simplified model is shown in Figure \ref{fig:ceRNA}A. Two species of mRNAs  are regulated by a common microRNA. Notice that the mRNAs do not directly interact in the model and all interactions are indirectly induced by the shared microRNA pool. We can represent this using a straight-forward generalization of the model considered earlier
\begin{align}
 & \frac{ds}{dt}=\alpha_{s}-\tau_{s}^{-1}s+\beta_1c_1+\beta_2c_2+\gamma_1c_1+\gamma_2c_2-(\mu_1 m_1+\mu_2 m_2)s+\ldots\nonumber\\
 &\qquad+\eta_{s}+\eta_{\beta_1}+\eta_{\beta_2}+\eta_{\gamma_1}+\eta_{\gamma_2}-\eta_{\mu_1}-\eta_{\mu_2}\\
 & \frac{dm_i}{dt}=\alpha_{m_i}-\tau_{m_i}^{-1}m_i+\gamma_i c_i-\mu_i m_is+\eta_{m_i}+\eta_{\gamma_i}-\eta_{\mu_i}\nonumber \\
 & \frac{dc_i}{dt}=\mu_i m_is-\beta_i c_i-\gamma_i c_i-\tau_{c_i}^{-1}c_i+\eta_{c_i}-\eta_{\beta_i}-\eta_{\gamma_i}+\eta_{\mu_i}\nonumber \\
 & \frac{dp}{dt}=\alpha_{p}m_1-\tau_{p}^{-1}p+\eta_{p}\nonumber 
\label{eqn:ceRNA}
\end{align}
with
\begin{align}
\overline{\eta_{j}(t)}&=0,\qquad j,k=s,\beta_i,\gamma_i,\mu_i,m_i,c_i,p\qquad ,i=1,2\\\nonumber
\overline{\eta_{j}(t)\eta_{k}(t+\tau)}&=\delta_{jk}\sigma_{j}^{2}\delta(\tau)
\end{align}
and  variances reflecting the Poisson nature of interactions: 
\begin{align}
\sigma_{s}^{2}&=\alpha_{s}+\tau_{s}^{-1}\overline{s},\quad\sigma_{\beta_i}^{2}=\beta_i \overline{c}_i,\quad\sigma_{\gamma_i}^{2}=\gamma_i \overline{c}_i,\quad\sigma_{\mu_i}^{2}=\mu_i \overline{m}_i\overline{s}\\\nonumber
\sigma_{m_i}^{2}&=\alpha_{m_i}+\tau_{m_i}^{-1}\overline{m}_i,\quad\sigma_{c_i}^{2}=\tau_{c_i}^{-1}\overline{c}_i,\quad\sigma_{p}^{2}=\alpha_{p}\overline{m}_1+\tau_{p}^{-1}\overline{p}\qquad ,i=1,2
\end{align}

The binding of microRNAs to mRNAs is controlled by the  interaction rates $\mu_{1,2}$. These rates reflect the binding affinity of microRNAs for the two mRNA species. We asked how protein noise and means change as we vary the  transcription rates,  $\alpha_{m_{1,2}}$, of the mRNAs. The remaining parameters are assumed to be the same for both mRNAs and from now on we have suppressed the indices on these parameters. MicroRNAs function catalytically and we focus on the parameter regime $\beta \gg \tau_c^{-1}, \gamma$.  Figure \ref{fig:ceRNA}B and C shows the mean protein levels, $\overline{p}_1$, and intrinsic noise of protein species $1$ as a function of the transcription rates  of the two mRNA genes, $\alpha_{m_{1,2}}$, for the case of equal binding affinities ($\mu_1= \mu_2$). Notice there is a peak in the noise similar to the single-species case. Once again there is good agreement between simulation and analytic calculations based on Langevin noise and the Linear Noise Approximation{ (see {\it Materials and Methods}).} We also examined the case where the mRNAs have different binding affinities for the microRNA, $\mu_1 = 0.2, \mu_2 = 2$. This results in an asymmetry in the noise peak but  does not change the major qualitative features of our results (see Figure \ref{fig:ceRNA}D and E)

As in the single-mRNA species case, the behavior of the system can be divided into regimes depending on whether the combined {normalized }transcription rate of both mRNA species is bigger or smaller than the sRNA transcription rates.  {We find the crossover regime to happen at ${\alpha_{m_1}\over\phi_1}+{\alpha_{m_2}\over\phi_2}\simeq\alpha_s$ with $\phi_i\equiv1+\beta_i\tau_{c_i} $(for derivation refer to } {\it Materials and Methods} {). This region} splits the transcription rate space $(\alpha_{m_1},\alpha_{m_2})$ into an expressing regime, {${\alpha_{m_1}\over\phi_1}+{\alpha_{m_2}\over\phi_2}\gtrsim\alpha_s$},  and a repressing regime, {${\alpha_{m_1}\over\phi_1}+{\alpha_{m_2}\over\phi_2}\lesssim\alpha_s$}. Here, similar to the single species case we see a sharp peak in the noise at the crossover regime ({Figure} \ref{fig:ceRNA}F). Since the crossover regime depends on {a linear combination of transcription rates, ${\alpha_{m_1}\over\phi_1}+{\alpha_{m_2}\over\phi_2}$},  { the existence of crosstalk can be determined by tracking the changes in crossover regime of one species as the transcription rate of another species is changed. In this regard measuring protein noise can be useful, since it has a peak that is easy to detect, whereas such a dramatic feature cannot be observed in the mean levels of proteins. This provides a tool for probing crosstalk between mRNAs in experimental setups if transcription rates can be tuned to access the crossover regime.}

\begin{figure}[!ht]
\begin{center}
\includegraphics[width=0.9\textwidth]{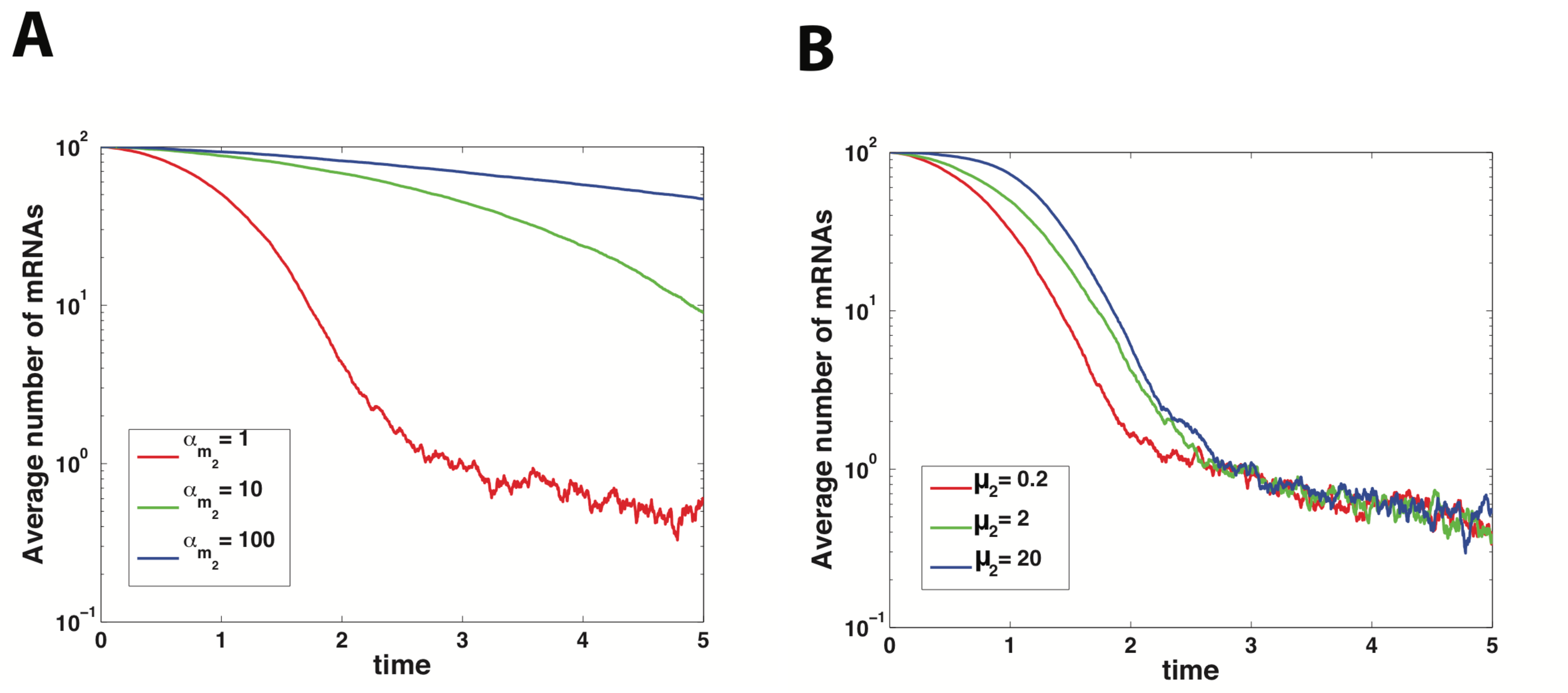}
\end{center}
\caption{{\textbf{mRNA Decay in ceRNA hypothesis.}} Time-course of mean mRNA number (first species) while transcription rate and interaction rate of the other species is being changed. Each curve is the average of 100 Gillespie simulations with the same initial conditions {at steady state of unregulated system ($m_1=\alpha_{m_1}\tau_{m_1}, m_2=\alpha_{m_2}\tau_{m_2}, s=0, c_1=c_2=0$)}. For both species parameters are the same as Figure \ref{fig:Intro} with $\alpha_{m_1}=\alpha_{m_2}=1, \mu_1=\mu_2=2, \beta=10$ with the exception of the parameter under study. \textbf{A)} Time course of mRNA as transcritption rate of the competing mRNA is changed. \textbf{B)} Time course of mRNA as interaction rate of the competing mRNA is changed. } 
\label{fig:TimeCourseCeRNA}
\end{figure}

{As an alternative experimental strategy for testing the ceRNA hypothesis, we studied mRNA decay in the context of ceRNA hypothesis, by tracking the mean mRNA number as a function of time before it reaches steady state. Figure  }\ref{fig:TimeCourseCeRNA}A {shows time-course of average mRNA number from species one as the transcription rate of mRNA species two is changed. Note that mRNA decays slower as the transcription rate of the competing mRNA is increased. This is due to access to a smaller pool of microRNAs, hence slower overall degradation. Figure }\ref{fig:TimeCourseCeRNA}B {shows the same plots as interaction rate of the competing mRNA with microRNA is increased. Again competing over shared pool of microRNA results in a slower degradation. This suggests an alternative experimental approach for determining crosstalk through noncoding RNAs based on mRNA decay. We propose that such crosstalk can be detected by changing any paramaters of an mRNA that increases its competing strength for microRNAs (e.g.  transcription rate, interaction rate) and probing changes in the decay rate of  other mRNAs.}

\sectionlarge{Discussion}
In this work, we studied gene regulation by {inhibitory} non-coding RNAs. Whereas gene regulation by prokaryotic sRNAs has been extensively studied \cite{Levine2007, Levine2008, Mehta2008, Mitarai2009, Jia2009, Platini2011, Baker2012}, there exist relatively few models of gene regulation by catalytic microRNAs \cite{Mukherji2011, Levine2007a, Hao2011}. Here, we used a simple kinetic model to study both the mean expression levels and intrinsic noise properties of post-transcriptional regulation by non-coding RNAs. Using a single parameter, our model interpolates between the stoichiometric behavior of prokaryotic sRNAs  and the catalytic behavior characteristic of eukaryotic microRNAs. We found that both sRNAs and microRNAs exhibit a  linear threshold behavior, with expressing and repressed regimes separated by a crossover regime. At the crossover, the mRNA transcription rate roughly equals the product of the non-coding-RNA transcription rate and the average number of mRNA molecules degraded by a single non-coding RNA molecule. In all cases, the intrinsic noise was smaller in the repressed regime and showed a sharp peak in the crossover regime. We found that for most parameter regimes, the intrinsic noise in the crossover regime shows an approximate data collapse, giving hints that there may be universal behavior near the transition from the repressed to expressing regime. We then generalized our calculations to study crosstalk between two mRNAs regulated by a single microRNA. We found that the intrinsic noise is an extremely sensitive measure of microRNA induced interactions between mRNAs. 

Our results for the mean expression profile is consistent with recent experimental studies \cite{Mukherji2011}. However, our conclusions about the intrinsic noise profiles of catalytic non-coding RNAs are different from {similar work done by} Hao  \emph{et al}. \cite{Hao2011}. They concluded that the intrinsic noise profiles of catalytic and stochiometric {interactions} differed significantly. The reason for this discrepancy is that Hao \emph{et al.} did not normalize the sRNA transcription rate $\alpha_s$ by $\phi$. Consequently, they compared the crossover regions of sRNA regulated genes to repressed regions of microRNAs. As shown above, after making this normalization there is extensive data collapse of intrinsic noise profiles for both stoichiometric and catalytic genes. 

One of the striking results of our calculation is the similarity between sRNA-regulated and microRNA-regulated genes. This similarity is somewhat surprising given that microRNAs and sRNAs are found in different kingdoms (prokaryotes versus eukaryotes) and utilize distinct biochemistry and molecular machinery. Eukaryotic microRNAs use  complicated nuclear machinery  to export microRNA into cytoplasm and bring it to mature state by incorporating the RNA strand into the RISC complex. In contrast, prokaryotic sRNAs undergo relatively little post-processing and bind mRNAs {spontaneously or} via the chaperone protein Hfq.  Given these extensive differences, the similarity between the expression characteristics of microRNA-regulated and sRNA regulated genes are suggestive of  convergent evolution. 

Our calculations show  that mRNAs regulated by catalytic non-coding RNAs have large peaks in the intrinsic noise. This differs significantly from results that would be derived from more traditional treatments of catalytic interactions based on the  Michaelis-Menten or Hill  equations. The underlying reason for this difference is that  traditionally, the Michaelis-Menten equations are derived under the assumption that substrates of enzymes are in excess compared to the enzymes themselves. In contrast, here we are interested in the case where the number of  enzymes (microRNAs) and number of substrates (mRNAs) are comparable. This accounts for the sharp peak in noise observed in the crossover regime in our model that is absent in traditional treatments of enzyme kinetics.

Our calculations also suggest a strategy for testing the ceRNA hypothesis \cite{Salmena2011}, which posits that microRNAs induce extensive interactions between mRNA molecules. Our calculations suggest that protein noise may be a sensitive measure of the competition between the two species. Thus, it may be possible to uncover interactions between mRNA by measuring changes in the noise of regulated genes. We suspect that this will be true even when we generalize our calculations to consider the case where $n$ mRNA species compete over the same pool of microRNAs. In this case, we hypothesize that there would still be a sharp peak in the intrinsic noise when the {total appropriately normalized transcription rates of all mRNAs equals the  transcription rate of microRNAs}. In the future, it will be interesting to analyze these more complicated models in greater detail.

{Finally we would like to note that the results discussed in this paper are based on a simplified version of the actual biochemical interactions with the goal of incorporating different mechanisms of gene regulation by noncoding RNAs into a simple mathematical framework. Adding further levels of complexity to the model can result in more realistic results, however the use of simple models can still provide insights into qualitative behavior of the system.  To determine the extent to which our results hold under more realistic biological assumptions further theoretical and experimental work is necessary.  The microRNAs discussed in this paper correspond to the RNA interference (RNAi) process and repress genes by binding to mRNAs. However, more recently, a small subset of microRNA processes, known as RNA activation (RNAa), has been discovered that can activate genes by binding to mRNAs} \cite{Henke2008,Niepmann2009,Truesdell2012}. The underlying mechanism for gene activation is not fully understood, but involves derepression of already repressed genes via microRNAs. In the future, it will be interesting to incorporate this biology into our model.
In addition,  mRNAs often undergo degradation in several stages. In this work,  we model  the degradation of mRNAs as a Poisson process with a single degradation constant. However it is well-known that the degradation of mRNAs is a complicated process often involving multiple steps \cite{Belasco2010,Huntzinger2011}.  Each additional step can be modeled by extra chemical interactions. These new steps do not affect the mean mRNA decay rates, and our results qualitatively match the experimental results\cite{Braun2011}. However these new interactions have additional noise that in general cannot be summed together as a single Poisson process. As a result our assumption of a single Poisson process for degradation underestimates noise. In this regard our results set a lower bound on the noise. 

{Finally, we note that prokaryotic and eukaryotic cells are likely to be described by very different parameter sets within our model.  In this paper we largely focused on changing the recycling ratio $\beta$ as a measure of catalytic versus stoichiometric interactions among mRNAs and noncoding RNAs. In doing so we were keeping the rates of production and decay of proteins and mRNAs constant. However these rates can vary significantly between prokaryotes and eukaryotes }\cite{Wang2002, Bernstein2002,Young1976,Schwanhausser2011,Vogel1994}. {Nevertheless, as shown in Figure} \ref{fig:ParameterChanges}, {our qualitative results are insensitive to most of the detailed kinetic parameters of regulation. This suggests our main conclusions should hold despite the differences in gene regulation between prokaryotes and eukaryotes.}

\sectionlarge{Materials and Methods}
\subsectionmedium{Gillespie Simulations and LNA}
We can convert between concentration and particle number by multiplying by a volume, $V$.  As is standard in the field, we have used notation where $V$ has been dropped from all the equations in the paper. The dependence on volume can be easily recovered by dimensional analysis. All Gillespie simulations (including ceRNA simulations) were done at the volume ($V=10$). The LNA results are also dependent on volume and are calculated with $V=10$.  As a result, throughout the paper, particle numbers from Gillepie simulations and concentrations from LNA have been accordingly scaled by volume to account for their correct dimensions. The choice of volume has been such that the problem is computationally and analytically tractable. For larger volumes the number of particles increases, leading to a decrease in fluctuations. As a result the approximate methods produce very accurate results. However the simulations become more computationally demanding. On the other hand, for smaller particle numbers LNA breaks down \cite{Grima2010,Grima2011}. It has been shown that in this case the peak in noise at crossover shifts from its expected theoretical value \cite{Mehta2008}, and even mean protein levels do not match with simulation. This discrepancy increases as volume (and hence particle number) is decreased. Hence the results presented in this paper will qualitatively hold for some smaller systems, but they cannot be generalized to arbitrarily small systems. 

\subsectionmedium{Single Species Linear noise approximation}
Linearization of equation \ref{eqn:SingleSpecies} results in:
\begin{align}
\frac{d}{dt}\begin{bmatrix} 
& \delta s\\
& \delta m\\
& \delta c\\
\end{bmatrix}=
\underbrace{\begin{bmatrix} 
&-\tau_{sR}^{-1} & -\mu \overline{s} & \beta+\gamma\\
&-\mu \overline{m} & -\tau_{mR}^{-1} & \gamma\\
&\mu \overline{m} & \mu \overline{s} & -\tau_{cR}^{-1}\\
\end{bmatrix}}_A
\begin{bmatrix} 
& \delta s\\
& \delta m\\
& \delta c\\
\end{bmatrix}+
\begin{bmatrix}
\eta_{s}+\eta_{\beta}+\eta_{\gamma}-\eta_{\mu}\\
\eta_{m}+\eta_{\gamma}-\eta_{\mu}\\
\eta_{c}-\eta_{\beta}-\eta_{\gamma}+\eta_{\mu}\\
\end{bmatrix}
\end{align}
where {$\overline{n}$ corresponds to the mean of species $n$ at steady state, $\delta n\equiv n-\overline{n}$ is the deviation from this value and for future reference} the transfer matrix {is named} $A$. {Furthermore we have defined the following effective lifetimes for $s, m$ and $c$}:
\begin{align}
\tau_{sR}^{-1}=\tau_{s}^{-1}+\mu \overline{m},\quad\tau_{mR}^{-1}=\tau_{m}^{-1}+\mu \overline{s},\quad \tau_{cR}^{-1}=\tau_c^{-1}+\beta+\gamma
\end{align}
To find the solution of this equation we apply a Fourier transform and solve the resulting equation:
\begin{align}
\begin{bmatrix} 
&\widetilde{\delta s}\\
&\widetilde{\delta m}\\
&\widetilde{\delta c}\\
\end{bmatrix}=
\begin{bmatrix} 
&i\omega+\tau_{sR}^{-1} & \mu \overline{s} & -\beta-\gamma\\
&\mu \overline{m} & i\omega+\tau_{mR}^{-1} & -\gamma\\
&-\mu \overline{m} & -\mu \overline{s} & i\omega+\tau_{cR}^{-1}\\
\end{bmatrix}^{-1}
\begin{bmatrix}
\tilde \eta_{s}+\tilde \eta_{\beta}+\tilde \eta_{\gamma}-\tilde \eta_{\mu}\\
\tilde\eta_{m}+\tilde\eta_{\gamma}-\tilde\eta_{\mu}\\
\tilde\eta_{c}-\tilde\eta_{\beta}-\tilde\eta_{\gamma}+\tilde\eta_{\mu}\\
\end{bmatrix}
\end{align}
where tilde denotes Fourier transform. So we have:
\begin{align}
\widetilde{\delta m}=\frac{1}{(i\omega-\lambda_1)(i\omega-\lambda_2)(i\omega-\lambda_3)}\begin{bmatrix} &F(\omega) &G(\omega) &H(\omega)\end{bmatrix}
\begin{bmatrix}
\tilde\eta_{s}+\tilde \eta_{\beta}+\tilde \eta_{\gamma}-\tilde \eta_{\mu}\\
\tilde\eta_{m}+\tilde\eta_{\gamma}-\tilde\eta_{\mu}\\
\tilde\eta_{c}-\tilde\eta_{\beta}-\tilde\eta_{\gamma}+\tilde\eta_{\mu}\\
\end{bmatrix}
\end{align}
where $\lambda_1,\lambda_2,\lambda_3$ are the eigenvalues of $A$ and:
\begin{align}
F(\omega)&=\gamma\mu \overline{m}-\mu \overline{m}(i\omega+\tau_{cR}^{-1})\\\nonumber
G(\omega)&=(i\omega+\tau_{sR}^{-1})(i\omega+\tau_{cR}^{-1})-(\beta+\gamma)\mu \overline{m}\\\nonumber
H(\omega)&=-(\beta+\gamma)\mu \overline{m}+\gamma(i\omega+\tau_{sR}^{-1})
\end{align}
or
\begin{align}
\widetilde{\delta m}=\frac{F(\omega)\tilde\eta_s+[F(\omega)-H(\omega)]\tilde\eta_\beta+[F(\omega)+G(\omega)-H(\omega)](\tilde\eta_\gamma-\tilde\eta_\mu)+G(\omega)\tilde\eta_m+H(\omega)\tilde\eta_c}
{(i\omega-\lambda_1)(i\omega-\lambda_2)(i\omega-\lambda_3)}
\end{align}
Since $\widetilde{\delta p}=\frac{\tilde\eta_{p}+\alpha_p\widetilde{\delta m}}{i\omega+\tau_{p}^{-1}}$, we have
\begin{align}
\overline{(\delta p)^2}&=\frac{\tau_p}{2}\sigma_p^2
+\int \frac{d\omega}{2\pi}\frac{\alpha_p^2}{\omega^2+\tau_p^{-2}}\times\\\nonumber
&\frac{\overbrace{|F(\omega)|^2\sigma^2_s+|F(\omega)-H(\omega)|^2\sigma^2_\beta+
|F(\omega)+G(\omega)-H(\omega)|^2(\sigma^2_\gamma+\sigma^2_\mu)
+|G(\omega)|^2\sigma^2_m+|H(\omega)|^2\sigma^2_c}^\text{\large{Q}}}
{(\omega^2+\lambda_1^2)(\omega^2+\lambda_2^2)(\omega^2+\lambda_3^2)}
\end{align}

We can rewrite Q as:
\begin{align}
Q&\equiv X\omega^4+Y\omega^2+Z
\end{align}

with 
\begin{align}
X&=(\gamma \tau_{cR}+1)\mu \overline{m} \,\overline{s}+\alpha_m+\tau_m^{-1}\overline{m}\\\nonumber
Y&= (\mu \overline{m})^2(\alpha_s+\tau_s^{-1}\overline{s})+(\mu \overline{m}+\gamma)^2\beta \tau_{cR}\mu \overline{m} \,\overline{s}\\\nonumber
&+(\tau_{s}^{-1}+\tau_{c}^{-1}+\beta)^2(\gamma \tau_{cR}+1)\mu \overline{m} \,\overline{s}\\\nonumber
&+[(\tau_{sR}^{-1}-\tau_{cR}^{-1})^2+2(\beta+\gamma)\mu \overline{m})^2](\alpha_m+\tau_m^{-1}\overline{m})+\gamma^2q\mu \overline{m} \,\overline{s}\\\nonumber
Z&=(\gamma-\tau_{cR}^{-1})^2(\mu \overline{m})^2(\alpha_s+\tau_s^{-1}\overline{s})
+(\tau_{c}^{-1}\mu \overline{m}+\gamma\tau_s^{-1})^2\beta \tau_{cR}\mu \overline{m} \,\overline{s}\\\nonumber
&+(\tau_{c}^{-1}+\beta)^2
(\gamma \tau_{cR}+1) \tau_{s}^{-2} \mu \overline{m} \,\overline{s}\\\nonumber
&+(\tau_{s}^{-1}\tau_{cR}^{-1}+\tau_{c}^{-1}\mu \overline{m})^2(\alpha_m+\tau_m^{-1}\overline{m})+(\beta\mu \overline{m}-\gamma\tau_{s}^{-1})^2q\mu \overline{m} \,\overline{s}\\\nonumber
\label{eqn:XYZ}
\end{align}
{ so we can write:} 
\begin{align}
\overline{(\delta p)^2}=\frac{\tau_p}{2}\sigma_p^2
+\alpha_p^2\int \frac{d\omega}{2\pi}\frac{X\omega^4+Y\omega^2+Z}
{(\omega^2+\lambda_1^2)(\omega^2+\lambda_2^2)(\omega^2+\lambda_3^2)(\omega^2+\lambda_4^{2})}
\end{align}

{where $\lambda_4\equiv-\tau_p^{-1}$. Now} by use of partial fractions and integration we get
\begin{align}
\overline{(\delta p)^2}&=\frac{\tau_p}{2}(\alpha_p \overline{m}+\tau_p^{-1}\overline{p})+
\sum_{i=1}^4\frac{\alpha_p^2}{2|\lambda_i|}(X\lambda_i^4-Y\lambda_i^2+Z)\prod_{j\not=i}\frac{1}{\lambda_j^2-\lambda_i^2}
\label{eqn:noise}
\end{align}

\begin{figure}[!ht]
\begin{center}
\includegraphics[width=0.9\textwidth]{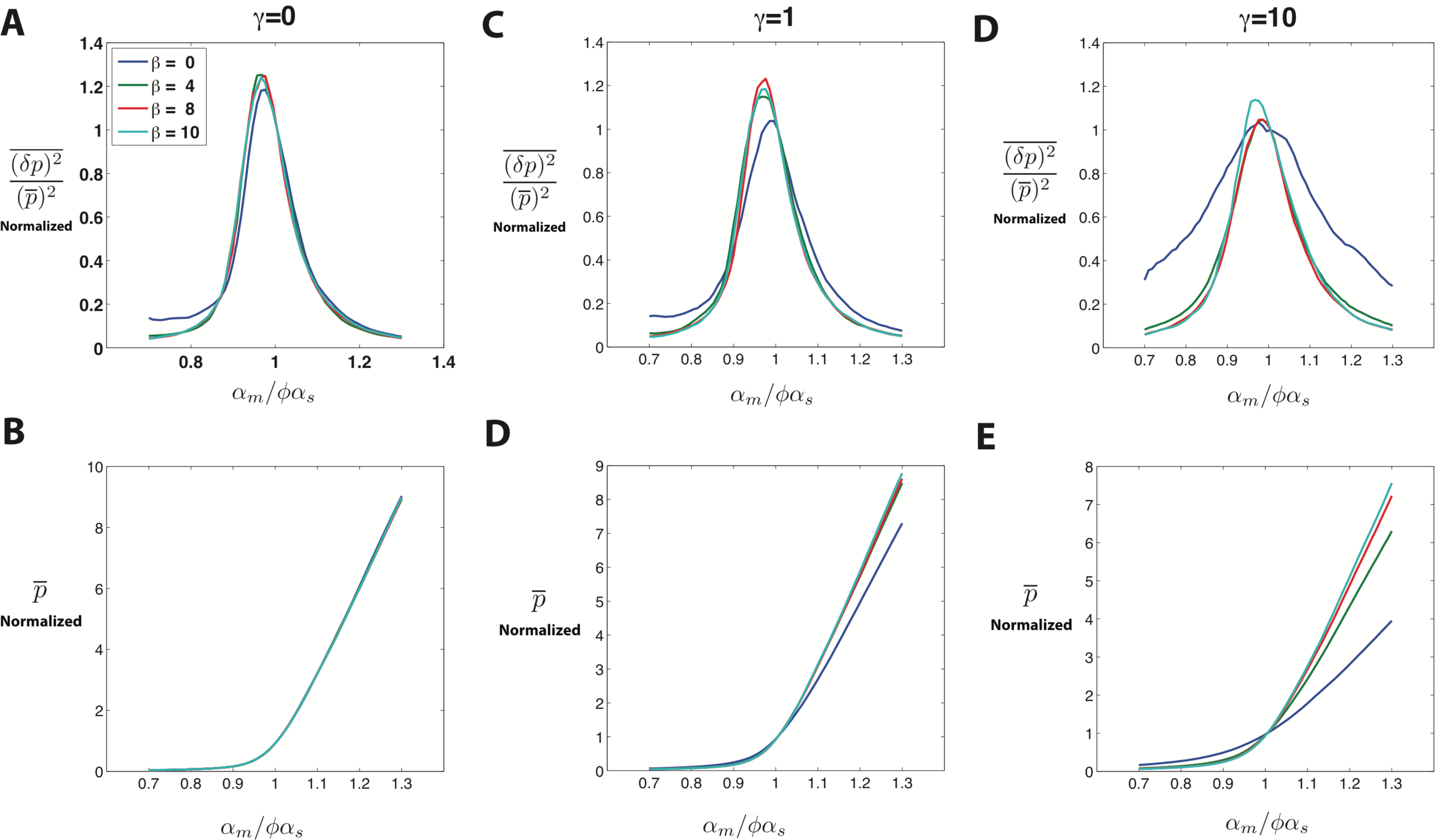}
\end{center}
\caption{{\textbf{Scaling at Crossover Regime (Simulation Results).}} Parameters same as Figure \ref{fig:Intro} with $\mu=2$. Data has been smoothed using a moving average method. {\bf A)} Protein noise normalized to its value at $\alpha_m=\phi\alpha_s$ (more explicitly ${\overline{(\delta p)^2}\over(\overline {p})^2} / \left[{\overline{(\delta p)^2}\over(\overline {p})^2}|_{{\alpha_m\over \phi\alpha_s}=1}\right]$) plotted as a function of $\alpha_m\over\phi\alpha_s$ for $\gamma=0$. Each line is a different value of $\beta$. Same legend for all figures. {\bf B)} Protein mean normalized to its value at $\alpha_m=\phi\alpha_s$ (more explicitly $\overline {p}/ \left[\overline {p}|_{{\alpha_m\over \phi\alpha_s}=1}\right]$) plotted as a function of $\alpha_m\over\phi\alpha_s$ for $\gamma=0$. {\bf C,D)} Graphs similar to  {\bf A,B} plotted for $\gamma=1$.  {\bf E,F)} Graphs similar to  {\bf A,B} plotted for $\gamma=10$. }
\label{fig:ScalingGillespie}
\end{figure}

\subsectionmedium{Single Species Scaling Results}
{As noted in the main text and Figure} \ref{fig:Scaling} {we see a scaling at the crossover regime using linear noise approximation. We see similar results using Gillespie algorithm as shown in Figure} \ref{fig:ScalingGillespie}. {The slight discrepancy between the two methods is partly due to the approximate nature of LNA and partly due to finite statistical sample size used in Gillespie. As a result the maximum in Gillespie algorithm is always slightly off from $\alpha_m=\phi\alpha_s$.}
\subsectionmedium{Single Species Asymptotic calculations}
\subsubsectionsmall{Expressing Regime }
In the expressing regime with large mRNA transcription rate we demand $\epsilon\equiv\frac{\lambda}{\alpha_m-\phi\alpha_s}\ll 1$. Expanding $\overline{m}$ and $\overline{s}$ in terms of $\epsilon$ we will get:
\begin{align}
\overline{m}= \frac{1}{q\mu\tau_s}\frac{1}{\epsilon} + O(\epsilon)\\\nonumber
\overline{s}=\alpha_s\tau_s\epsilon+O(\epsilon^2)\\
\label{eqn:MeanExpressing}
\end{align}
To find the eigenvalues of the transfer matrix $A$ we will expand it in terms of $\epsilon$:
\begin{align}
A
=\begin{bmatrix} 
&-\tau_{s}^{-1}-\tau_{s}^{-1}q^{-1}\epsilon^{-1}+O(\epsilon) &  -\mu \alpha_s\tau_s\epsilon+O(\epsilon^2)& \beta+\gamma\\
&-\tau_s^{-1}q^{-1}\epsilon^{-1}+O(\epsilon)& -\tau_{m}^{-1} -\mu \alpha_s\tau_s\epsilon+O(\epsilon^2)& \gamma\\
&\tau_s^{-1}q^{-1}\epsilon^{-1}+O(\epsilon)  &  \mu \alpha_s\tau_s\epsilon+O(\epsilon^2) & -\tau_{cR}^{-1}\\
\end{bmatrix}
\end{align}

the eigenvalues of this matrix satisfy the following equation:
\begin{align}
\lambda^3+a_2\lambda^2
+a_1\lambda+a_0=0
\end{align}

where
\begin{align}
a_2&=\tau_{s}^{-1}q^{-1}\epsilon^{-1}+O(1)\\\nonumber
a_1&=\tau_s^{-1}q^{-1}(\tau_m^{-1}+\tau_{c}^{-1})\epsilon^{-1}+O(1)\\\nonumber
a_0&=q^{-1}\tau_s^{-1}\tau_m^{-1}\tau_c^{-1}\epsilon^{-1}+O(1)
\end{align}
this equation can be analytically solved by expanding $\lambda$'s in terms of $\epsilon$, and results in the following {eigenvalues}:
\begin{align}
\lambda_1&=-\tau_m^{-1} + O(\epsilon)\\\nonumber
\lambda_2&=-\tau_c^{-1} + O(\epsilon)\\\nonumber
\lambda_3&=-\frac{\tau_s^{-1}q^{-1}}{\epsilon}  + O(1) 
\end{align}
{with} one fast mode and two slow modes. Next we will calculate the noise by expanding equation \ref{eqn:XYZ} in terms of $\epsilon$, which results in (making the substitution $C\equiv\tau_s^{-1}q^{-1}$)
\begin{align}
X&\simeq\frac{2\lambda}{\epsilon}\\\nonumber
Y&\simeq \frac{2\lambda}{q^2\tau_s^2\epsilon^3}=\frac{C^2X}{\epsilon^2} \\\nonumber
Z&\simeq\frac{2\lambda}{q^2\tau_s^2\tau_c^2\epsilon^3}=\frac{\tau_c^{-2}C^2X}{\epsilon^2}\\\nonumber
\end{align}

plugging this result into equation \ref{eqn:noise} gives:
\begin{align}
\overline{(\delta p)^2}&=\overline{p}+
\sum_{i=1}^4\frac{\alpha_p^2 \lambda}{\epsilon |\lambda_i|}(\lambda_i^4-C^2\epsilon^{-2}\lambda_i^2+C^2\epsilon^{-2}\tau_c^{-2})\prod_{j\not=i}\frac{1}{\lambda_j^2-\lambda_i^2}
\end{align}

{Now make explicit substitutions for $\lambda_i$ and only keep highest order in $\epsilon$ to get:}
\begin{align}
\overline{(\delta p)^2}
&\simeq\overline{p}
+\frac{\alpha_p^2 \lambda}{\epsilon \tau_p^{-1}}\frac{(\tau_c^{-2}-\tau_p^{-2})C^2\epsilon^{-2}}
{\left(\tau_m^{-2}-\tau_p^{-2}\right)\left(\tau_c^{-2}-\tau_p^{-2}\right)C^2\epsilon^{-2}}
+\frac{\alpha_p^2 \lambda}{\epsilon \tau_m^{-1}}\frac{(\tau_c^{-2}-\tau_m^{-2})C^2\epsilon^{-2}}
{\left(\tau_p^{-2}-\tau_m^{-2}\right)\left(\tau_c^{-2}-\tau_m^{-2}\right)C^2\epsilon^{-2}}\\\nonumber
&+\frac{\alpha_p^2 \lambda}{\epsilon \tau_c^{-1}}\frac{\tau_c^{-4}}
{\left(\tau_p^{-2}-\tau_c^{-2}\right)\left(\tau_m^{-2}-\tau_c^{-2}\right)C^2\epsilon^{-2}}
+\alpha_p^2 \lambda C^{-1}\frac{\tau_c^{-2}C^{2}\epsilon^{-2}}{C^6\epsilon^{-6}}\\\nonumber
\end{align}

{The terms on the first line above are order $\epsilon^{-1}$, while the terms on the second line are higher order in $\epsilon$ and can be ignored (order $\epsilon$ and $\epsilon^4$ respectively). Further simplification leads to: }

\begin{align}
\overline{(\delta p)^2}
 &\simeq \overline{p}+\alpha_p^2\lambda\epsilon^{-1}\left(\frac{\tau_p}{\tau_m^{-2}-\tau_p^{-2}}+\frac{\tau_m}{\tau_p^{-2}-\tau_m^{-2}} \right)\\\nonumber
&\simeq \overline{p}+\alpha_p^2\lambda\epsilon^{-1}\left(\frac{\tau_p^2\tau_m^2}{\tau_m+\tau_p} \right)
\simeq \overline{p}+{\alpha_p^2}\alpha_m \left(\frac{\tau_p^2\tau_m^2}{\tau_m+\tau_p} \right)=\overline{p}\left(1+b\frac{\tau_p}{\tau_p+\tau_m}\right)\\\nonumber
\end{align}

{where in the last line it is assumed that in the expressing regime $\alpha_m\gg\phi\alpha_s$, hence $\epsilon\simeq {\lambda\over\alpha_m}$.} {Dividing both sides of this equation by $\overline{p}^2$ and using the assumption $\tau_p\gg\tau_m$ gives the final result:}
\begin{align}
\frac{\overline{(\delta p)^2}}{\overline{p}^2}\simeq \frac{1+b}{\overline{p}}
\end{align}

\subsubsectionsmall{Repressing Regime}
In this regime with mRNA transcription very small, we demand $\epsilon\equiv\frac{\lambda}{\phi\alpha_s-\alpha_m}\ll 1$. Expansion of $\overline{m}$ and $\overline{s}$ in terms of $\epsilon$ gives:

\begin{equation}
\begin{split}
\overline{m}=\alpha_m\tau_m\epsilon+O(\epsilon^2),\qquad
\overline{s}=\frac{1}{q\mu\tau_m\phi\epsilon}+O(\epsilon)
\end{split}
\label{eqn:MeanRepressed}
\end{equation}

Next we expand the transfer matrix $A$ in terms of $\epsilon$:
\begin{align}
A
=\begin{bmatrix} 
&-\tau_{s}^{-1}-\mu\alpha_m\tau_m\epsilon+O(\epsilon^2) & -\frac{1}{q\tau_m\phi\epsilon}+O(\epsilon)& \beta+\gamma\\
&-\mu\alpha_m\tau_m\epsilon+O(\epsilon^2) & -\tau_{m}^{-1} -\frac{1}{q\tau_m\phi\epsilon}+O(\epsilon)& \gamma\\
&\mu\alpha_m\tau_m\epsilon+O(\epsilon^2)& \frac{1}{q\tau_m\phi\epsilon}+O(\epsilon)& -\tau_{cR}^{-1}\\
\end{bmatrix}
\end{align}
after some calculation we find the following closed form for eigenvalues of $A$:
\begin{align}
\lambda^3+a_2\lambda^2
+a_1\lambda+a_0=0
\end{align}

where
\begin{align}
a_2&=\tau_{m}^{-1}q^{-1}\phi^{-1}\epsilon^{-1}+O(1)\\\nonumber
a_1&=\tau_{m}^{-1}q^{-1}\phi^{-1}(\tau_s^{-1}+\tau_{c}^{-1}+\beta)\epsilon^{-1}+O(1)\\\nonumber
a_0&=\tau_{m}^{-1}q^{-1}\phi^{-1}\tau_s^{-1}(\tau_c^{-1}+\beta)\epsilon^{-1}+O(1)
\end{align}
this equation can be analytically solved by expanding $\lambda$'s in terms of $\epsilon$, and results in the following:
\begin{align}
\lambda_1&=-\tau_s^{-1}+O(\epsilon) \\\nonumber
\lambda_2&=-(\tau_c^{-1}+\beta)+O(\epsilon) \\\nonumber
\lambda_3&=-\frac{\tau_m^{-1}q^{-1}\phi^{-1}}{\epsilon}  +O(1)
\end{align}
which has one fast mode and two slow modes. Next we calculate the noise by expanding equation \ref{eqn:XYZ} in terms of $\epsilon$: 
\begin{align}
X&\simeq 2\alpha_m q^{-1} \phi^{-1}\\\nonumber
Y&\simeq 2\alpha_m\left(\tau_s^{-2}\tau_{cR}^{-1}q^{-1}\phi^{-1}+\gamma \tau_s^{-1}+\tau_{cR}^{-2}\right)\\\nonumber
Z&\simeq 2\alpha_m\tau_s^{-2}\tau_{cR}^{-2}\nonumber
\end{align}
plugging this result into equation \ref{eqn:noise} gives:

\begin{align}
\frac{\overline{(\delta p)^2}}{\overline{p}^2}
&\simeq \frac{1}{\overline{p}}+
\frac{b\epsilon}{\overline{p}}\frac{q^2\phi^2}{2\alpha_m\tau_p}\Bigg(\frac{X\tau_p^{-4}-Y\tau_p^{-2}+Z}
{\left(\tau_s^{-2}-\tau_p^{-2}\right)\left((\tau_c^{-1}+\beta)^2-\tau_p^{-2}\right)\tau_p^{-1}}
+\frac{X\tau_s^{-4}-Y\tau_s^{-2}+Z}
{\left(\tau_p^{-2}-\tau_s^{-2}\right)\left((\tau_c^{-1}+\beta)^2-\tau_s^{-2}\right)\tau_s^{-1}}\\\nonumber
&\qquad\qquad+\frac{X(\tau_c^{-1}+\beta)^4-Y(\tau_c^{-1}+\beta)^2+Z}
{\left(\tau_p^{-2}-(\tau_c^{-1}+\beta)^2\right)\left(\tau_s^{-2}-(\tau_c^{-1}+\beta)^2\right)(\tau_c^{-1}+\beta)}\Bigg)
+O(\epsilon^3)
\end{align}
in the main text we have used the shorthand for the second term on the right hand side such that $\frac{\overline{(\delta p)^2}}{\overline{p}^2}\equiv {1+b\zeta\epsilon\over\overline{p}}$. For $\beta\gg \gamma, \tau_c^{-1}, \tau_s^{-1}, \tau_p^{-1}, $ after some calculation we get $\zeta\simeq 1$ leading to:
\begin{align}
\frac{\overline{(\delta p)^2}}{\overline{p}^2}\simeq {1+b\epsilon\over\overline{p}}
\end{align}

\subsectionmedium{ceRNA Linear Noise Approximation}
For two mRNAs, we linearized equation \ref{eqn:ceRNA} as:

\begin{align}
\frac{d\delta\chi}{dt}=A\delta\chi+\xi
\end{align}

\begin{align}
\delta\chi&=\begin{bmatrix} 
& \delta s\\
& \delta m_1\\
& \delta m_2\\
& \delta c_1\\
& \delta c_2\\
& \delta p_1\\
& \delta p_2\\
\end{bmatrix},\quad
\xi=\begin{bmatrix}
\xi_1\\
\xi_2\\
\xi_3\\
\xi_4\\
\xi_5\\
\xi_6\\
\xi_7\\
\end{bmatrix},\\\nonumber
A&=\begin{bmatrix} 
&-\tau_{sR}^{-1} & -\mu_1 \overline{s} & -\mu_2 \overline{s} & \beta_1+\gamma_1 & \beta_2+\gamma_2 & 0 & 0\\
&-\mu_1 \overline{m}_1 & -\tau_{mR_1}^{-1} & 0 & \gamma_1 & 0 & 0 & 0\\
&-\mu_2 \overline{m}_2 & 0 & -\tau_{mR_2}^{-1} & 0 & \gamma_2  & 0 & 0\\
& \mu_1 \overline{m}_1 & \mu_1 \overline{s} & 0 & -\tau_{cR_1}^{-1} & 0 & 0 & 0\\
& \mu_2 \overline{m}_2 & 0 &  \mu_2 \overline{s} & 0 & -\tau_{cR_2}^{-1} & 0 & 0 \\
&  0 & \alpha_{p_1} & 0 & 0 & 0 & -\tau_{p_1}^{-1} & 0\\
&  0 & 0 & \alpha_{p_2} & 0 & 0 & 0 & -\tau_{p_2}^{-1}\\
\end{bmatrix}
\end{align}

with 
\begin{align}
\tau_{sR}^{-1}=\tau_s^{-1}+\mu_1\overline{m}_1+\mu_2\overline{m}_2,\qquad \tau_{mR_i}^{-1}=\tau_{m_i}^{-1}+\mu_i\overline{s},\qquad \tau_{cR_i}^{-1}=\tau_{c_i}^{-1}+\gamma_i+\beta_i
\end{align}
and:
\begin{align}
\overline{\xi_i(t)\xi_j(t+\tau)}=\Gamma_{ij}\delta(\tau)
\end{align}
where $\Gamma_{ij}$'s are the elements of the following matrix:
\begin{align}
\Gamma=\begin{bmatrix}
g_1 & g_2 & g_3 & g_4 & g_5  & 0 & 0 \\
g_2 & g_6 & 0     & g_7 & 0   &  0 & 0 \\
g_3 & 0 & g_8 & 0 & g_9 & 0 &0 \\
g_4  & g_7 & 0 & g_{10} & 0 & 0 & 0 \\
g_5 & 0 & g_9 & 0 & g_{11} & 0 & 0\\
0 & 0 & 0 & 0 & 0 & g_{12} & 0\\
0 & 0 & 0 & 0 & 0 & 0 & g_{13}\\
\end{bmatrix}
\end{align}

and 
\begin{align}
g_1&=\sigma_s^2+\sigma_{\beta_1}^2+\sigma_{\gamma_1}^2-\sigma_{\mu_1}^2+\sigma_{\beta_2}^2+\sigma_{\gamma_2}^2-\sigma_{\mu_2}^2\\\nonumber
g_2&=\sigma_{\gamma_1}^2+\sigma_{\mu_1}^2\\\nonumber
g_3&=\sigma_{\gamma_2}^2+\sigma_{\mu_2}^2\\\nonumber
g_4&=-\sigma_{\beta_1}^2-\sigma_{\gamma_1}^2-\sigma_{\mu_1}^2\\\nonumber
g_5&=-\sigma_{\beta_2}^2-\sigma_{\gamma_2}^2-\sigma_{\mu_2}^2\\\nonumber
g_6&=\sigma_{m_1}^2+\sigma_{\gamma_1}^2+\sigma_{\mu_1}^2\\\nonumber
g_7&=-\sigma_{\gamma_1}^2-\sigma_{\mu_1}^2\\\nonumber
g_8&=\sigma_{m_2}^2+\sigma_{\gamma_2}^2+\sigma_{\mu_2}^2\\\nonumber
g_9&=-\sigma_{\gamma_2}^2-\sigma_{\mu_2}^2\\\nonumber
g_{10}&=\sigma_{c_1}^2+\sigma_{\beta_1}^2+\sigma_{\gamma_1}^2+\sigma_{\mu_1}^2\\\nonumber
g_{11}&=\sigma_{c_2}^2+\sigma_{\beta_2}^2+\sigma_{\gamma_2}^2+\sigma_{\mu_2}^2\\\nonumber
g_{12}&=\sigma_{p_1}^2\\\nonumber
g_{13}&=\sigma_{p_2}^2\\\nonumber
\end{align}

\begin{figure}[!ht]
\begin{center}
\includegraphics[width=0.9\textwidth]{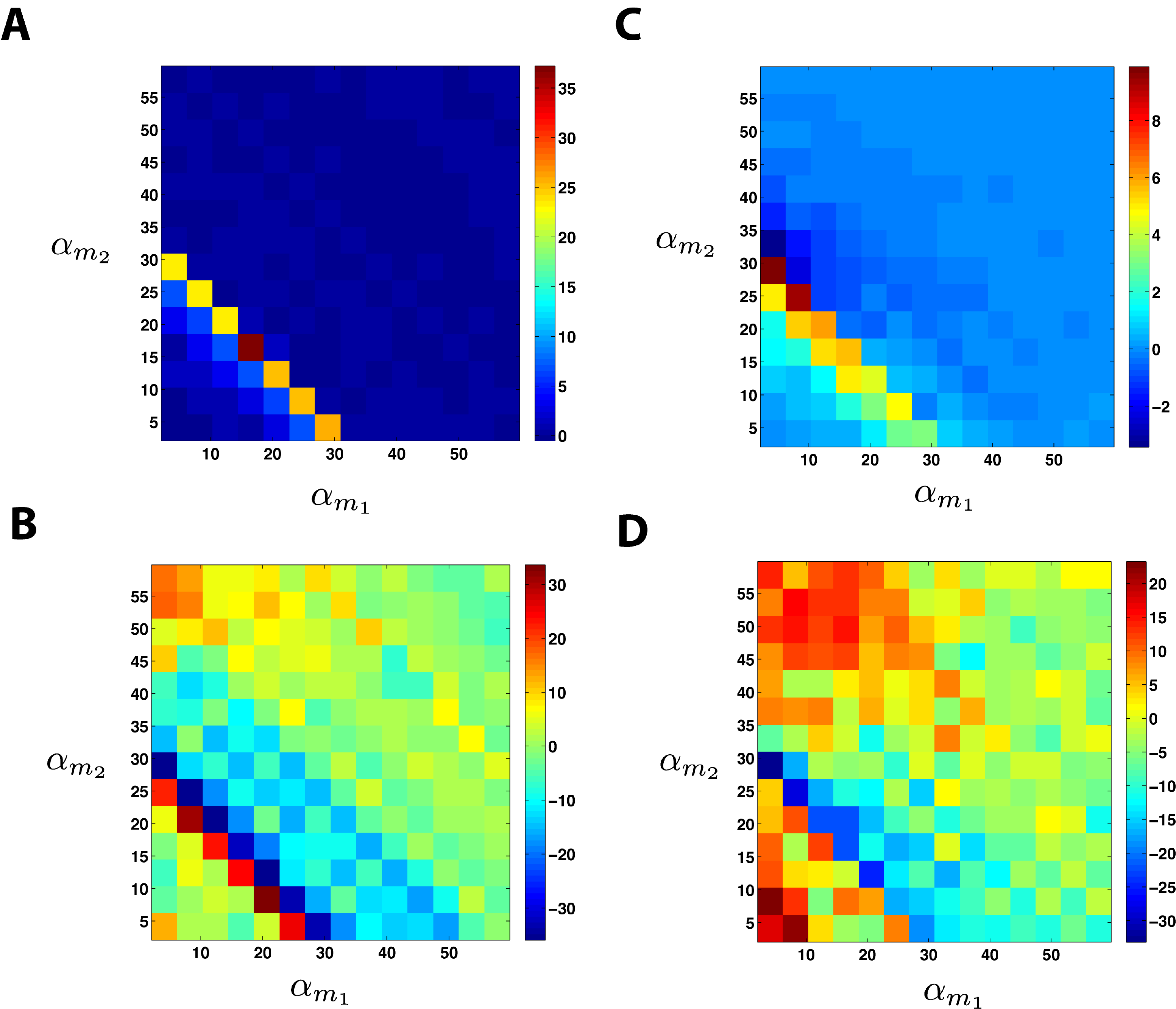}
\end{center}
\caption{\textbf{ceRNA Error.} { Figure showing the percentage of difference between LNA results and Gillespie results in the ceRNA hypothesis calculated as $error={Gillespie-LNA\over LNA}\times 100\%$} \textbf{A,B)} Error of protein mean (A) and noise (B) as a function of transcription rates of the two mRNAs with equal $\mu's$. Parameters same as in Figure \ref{fig:ceRNA}B. \textbf{C,D)} Error of protein mean (C) and noise (D) as a function of transcription rates of the two mRNAs with unequal $\mu's$. Parameters same as in Figure \ref{fig:ceRNA}D.}
\label{fig:ceRNAErrors}
\end{figure}

We find corresponding two point correlation functions by use of the following equation \cite{Swain2004}:
\begin{align}
\overline{\delta\chi_i\delta\chi_j}=-\sum_{p,q,r,s}B_{ip}B_{jr}\frac{\Gamma_{qs}}{\lambda_p+\lambda_r}B^{-1}_{pq}B^{-1}_{rs}
\end{align}

where $\lambda$'s  are the eigenvalues of $A$, and $B$ is the matrix of eigenvectors, according to:
\begin{align}
\sum_{j}A_{ij}B_{ij}=\lambda_kB_{ik}
\end{align}

we saw a good agreement between these results and Gillespie simulations. The two methods showed at most a deviation of 30\% from each other {(see Figure} \ref{fig:ceRNAErrors} {for more information)}.

\subsectionmedium{ceRNA Asymptotics}
For the steady state values of concentrations we have:
\begin{align}
\overline{c}_i&=\frac{\mu_i\overline{m}_i\overline{s}}{\tau_{cR_i}^{-1}}\\
\overline{m}_i&=\frac{\alpha_{m_i}}{\tau_{m_i}^{-1}+q_i\phi_i\mu_i\overline{s}}\\
\alpha_s&=\tau_s^{-1}\overline{s}-\sum_{i=1}^2(\beta_i+\gamma_i)\frac{\mu_i\overline{m}_i}{\tau_{cR_i}^{-1}}\overline{s}+\sum_{i=1}^2\mu_i\overline{m}_i\overline{s}
\end{align}
which results in 
\begin{align}
\alpha_s=\tau_s^{-1}\overline{s}+\sum_{i=1}^2 \frac{q_i\alpha_{m_i}\mu_i}{\tau_{m_i}^{-1}+q_i\phi_i\mu_i\overline{s}}\overline{s}
\end{align}
After some calculations we derive the following polynomial for sRNA concentration, $\overline{s}$:
\begin{align}
\overline{s}^3+(B_1+B_2+A_1+A_2-K)\overline{s}^2+(B_1B_2-K(B_1+B_2)+A_1B_2+A_2B_1)\overline{s}-KB_1B_2=0
\end{align}
with 
\begin{align}
K=\alpha_s\tau_s,\qquad A_i=\frac{\tau_s\alpha_{m_i}}{\phi_i},\qquad B_i=\frac{\tau_s\lambda_i}{\phi_i}, \qquad\lambda_i=\frac{1}{q_i\mu_i\tau_{m_i}\tau_s}
\end{align}
Furthermore, for ease of notation, in what follows we will use the following definitions: 
\begin{align}
A_T=A_1+A_2,\quad B_T=B_1+B_2
\end{align}
\subsubsectionsmall{Expressing Regime}
In this regime $A_T\gg G\equiv max(B_T,KB_T,B_1B_2)$ and we can simplify the polynomial equation by defining $\epsilon\equiv G/A_T$ and multiplying it by $\epsilon$ while keeping coeffiecients to first order:
\begin{align}
\epsilon \overline{s}^3+G\overline{s}^2+GD\overline{s}-KB_1B_2\epsilon=0
\end{align}
with $D\equiv\frac{A_1}{A_T}B_2+\frac{A_2}{A_T}B_1$. Note that $GD$ is of order $O(\epsilon^0)$ and does not require $\epsilon$ expansion. The equation for $\overline{s}$ has the following asymptotic solutions:
\begin{align}
\overline{s}&=-\frac{G}{\epsilon},-D, \frac{kB_1B_2}{GD}\epsilon
\end{align}
with the only positive solution being $\overline{s}=\frac{kB_1B_2}{GD}\epsilon=\frac{\alpha_s\tau_s}{\frac{\alpha_{m_1}}{\lambda_1}+\frac{\alpha_{m_2}}{\lambda_2}}$. In the limit of $\alpha_{m_2}=0$, this  reduces to the single species result in equation \ref{eqn:MeanExpressing}. Finally for mRNA we have $\overline{m_i}\simeq \tau_{m_i}\alpha_{m_i}(1-q_i\phi_i\tau_{m_i}\mu_i\overline{s})\simeq\tau_{m_i}\alpha_{m_i}+O(\epsilon)$ which is the expected result in the expressing regime.

\subsubsectionsmall{Repressing Regime}
In this regime $K\gg G\equiv max(A_i,B_i)$ which is equivalent to $\alpha_s\gg{\alpha_{m_i}\over\phi_i},{\lambda_i\over\phi_i}$. Using this assumption we can simplify the polynomial equation by defining $\epsilon=G/K$ and multiplying it by $\epsilon$ while keeping coeffiecients to first order:
\begin{align}
 \epsilon \overline{s}^3-(G-(A_T+B_T)\epsilon)\overline{s}^2+((B_1B_2+A_1B_2+A_2B_1)\epsilon-GB_T)\overline{s}-GB_1B_2=0
\end{align}
which has the following asymptotic solutions:
\begin{align}
\overline{s}=-B_1,\quad -B_2,\quad \frac{G}{\epsilon}+x
\end{align}
with $x$ being the solution for the following equation:
\begin{align}
\epsilon(\frac{G}{\epsilon}+x)^3-(G-(A_T+B_T)\epsilon)(\frac{G}{\epsilon}+x)^2-GB_T\frac{G}{\epsilon}=0
\end{align}
This results in $x=-A_T$. So the only positive solution is $\overline{s}\simeq\frac{G}{\epsilon}-A_T=\tau_s(\alpha_s-{\alpha_{m_1}\over\phi_1}-{\alpha_{m_2}\over\phi_2})$ and $\overline{m}_i\simeq \frac{\alpha_{m_i}\tau_{m_i}\lambda_i\phi_i^{-1}}{\alpha_s-{\alpha_{m_1}\over\phi_1}-{\alpha_{m_2}\over\phi_2}}$ which for $\alpha_{m_2}=0$ reduces to our single species results in equation \ref{eqn:MeanRepressed}.

\subsubsectionsmall{Crossover Regime}
{Crossover regime is roughly where the asymptotic solution of mean noncoding RNA in the repressed regime, $\overline{s}\simeq\tau_s(\alpha_s-{\alpha_{m_1}\over\phi_1}-{\alpha_{m_2}\over\phi_2})$, intersects with $\overline{s}=0$, hence at this point we have ${\alpha_{m_1}\over\phi_1}+{\alpha_{m_2}\over\phi_2}\simeq\alpha_s$}

\sectionlarge{Acknowledgments}
\bibliography{references.bib}

\sectionlarge{Figure Legends}

\end{document}